\documentclass[journal]{IEEEtran}
\usepackage{bm}
\usepackage{amsfonts}
\usepackage{amsmath}
\usepackage{color}

\usepackage{graphicx}  
\usepackage{epstopdf}
\usepackage{multirow}
\usepackage{booktabs}

\begin{document}

\title{Early Anomaly Detection and {Localization} in Distribution Network: A Data-Driven Approach}

\author{ Xin Shi$^1$,~\IEEEmembership{Student Member,~IEEE}, Robert Qiu$^{1,2}$,~\IEEEmembership{Fellow,~IEEE}, Xing He$^1$,~\IEEEmembership{Member,~IEEE}, \\
Zenan Ling$^1$, Haosen Yang$^1$, Lei Chu$^1$,~\IEEEmembership{Student Member,~IEEE}
\thanks{This work was partly supported by NSF of China No. 61571296, National Key R \& D Program of No. 2018YFF0214705  and (US) NSF Grant No. CNS-1619250.

$^1$ Department of Electrical Engineering, Center for Big Data and Artificial Intelligence, Shanghai Jiaotong University, Shanghai 200240, China.(e-mail: dugushixin@sjtu.edu.cn; rcqiu@sjtu.edu.cn; hexing\_hx@126.com; ling\_zenan@163.com; 2457604987@qq.com; leochu@sjtu.edu.cn;)

$^2$ Department of Electrical and Computer Engineering,Tennessee Technological University, Cookeville, TN 38505, USA. (e-mail:rqiu@tntech.edu)

}
}

\maketitle

\begin{abstract}
The measurement data collected from the supervisory control and data acquisition (SCADA) system installed in distribution network can reflect the operational state of the network effectively. In this paper, a random matrix theory (RMT) based approach is developed for early anomaly detection and localization by using the data. For every feeder in the distribution network, a corresponding data matrix is formed. Based on the Marchenko-Pastur Law for the empirical spectral analysis of covariance `signal+noise' matrix, the linear eigenvalue statistics are introduced to indicate the anomaly, and the outliers and their corresponding eigenvectors are analyzed for locating the anomaly. As for the low observability feeders in the distribution network, an increasing data dimension algorithm is designed for the formulated low-dimensional matrices being more accurately analyzed. The developed approach can detect and localize the anomaly at an early stage, and it is robust to random disturbance and measurement error. Cases on Matpower simulation data and real SCADA data corroborate the feasibility of the approach.
\end{abstract}
\begin{IEEEkeywords}
early anomaly detection and {localization}, distribution network, {SCADA} data, random matrix theory, increasing data dimension
\end{IEEEkeywords}

\IEEEpeerreviewmaketitle

\section{Introduction}
\label{section: Introduction}

\IEEEPARstart{T}{he} distribution network is an important part of the power system, and its operating state is directly related to the safety of the entire system. One main factor that influences the operating state of the distribution network are the anomalies caused by overload, unbalanced three-phase voltage or current, system swing, etc. The anomalies, in general, may last for a period of time, but if can't be detected and located in time, they will be likely to expand and even cause power failures or system black out. {Therefore, it is important to realize anomaly detection and localization at an early stage for the safety analysis and decision making of control strategy. In practice, the anomaly usually generates complex, nonliear and  intermittent features with random magnitue \cite{jaafari2007underground}, which makes it difficult for detecting and localizing them at an early stage. Meanwhile, with the increasing expansion of distribution network, it becomes more difficult for the model-based approaches to realize early anomaly detection and localization for the numerous branch lines and complex network topology.} 

Along the years, there have been significant deployments of online monitoring devices in power systems, which lies the foundation to enable a true monitoring, such as linear state estimation \cite{Sarri2016Performance}, dynamic state estimator \cite{Aminifar2014Power} or fully measurements of all state variables \cite{alcaide2018electric}. {The massive data collected from them can reflect the operational states of the system effectively, which stimulates the researches on data analytics for anomaly detection and {localization. In \cite{ma2003time}, a novelty detection approach based on one-class support vector machines (SVMs) is developed for time-series data analysis.} In \cite{xie2014dimensionality}, based on principal component analysis (PCA), the dimensionality of the phasor-measurement-unit (PMU) data is reduced for detecting the anomaly events in power systems at an eraly stage. {In \cite{malhotra2015long}, a long short term memory (LSTM) network based approach is proposed for analyzing the time-series data.} In \cite{Chu2016Massive}, the PMU data is modeled and analyzed by applying the hypothesis test theory for multiple covariance matrix. In \cite{pignati2017fault}, a real-time anomaly detection and abnormal line identification approach is developed, which merges the early anomaly detection and {localization} functionalities. In \cite{wu2017online}, a density-based detection algorithm is proposed to detect local outliers, which can differentiate high-quality synchrophasor data from the low-quality one during system physical disturbance. {In \cite{liu2018anomaly}, a structured autoencoder network is designed for detecting abnormal behavior in manufacture systems.}

Due to the massive data collected in a distribution network, the demand for theories capable of processing high-dimensional data has grown dramatically. The random matrix theory (RMT), introduced by Wishart in 1928 \cite{wishart1928generalised}, is an important mathematical tool for statistical analysis of high-dimensional data.
As for high-dimensional random matrices, the importance of the RMT for statistics comes from the fact that it may be used to correct traditional tests or estimators which fail in the `large $p$, large $n$' setting, where $p$ is the number of parameters (dimensions) and $n$ is the sample size. The RMT starts with asymptotic theorems on the distribution of eigenvalues or singular values of random matrices with certain assumptions, and eventually gives macroscopic quantity to indicate the data behavior. The theorems ensure the convergence of the empirical eigenvalue distributions to deterministic functions as the matrices grow large, which makes the RMT naturally suitable for high-dimensional data analysis. Nowadays, the RMT has been widely used in wireless communication \cite{qiu2012cognitive}, finance \cite{saad2013random}, quantum information \cite{chaitanya2015random}, etc. In recent years, some work that makes substantial use of results in the RMT has emerged in the power field. For example, in \cite{he2017big}, an architecture with the application of the RMT into smart grid is proposed. In \cite{xu2017correlation}, based on the RMT, a data-driven approach to reveal the correlations between various factors and the power system status is proposed. In \cite{Liuwei2016} and \cite{Wuxi2016}, the RMT is used for power system transient analysis and steady-state analysis, respectively.

In this paper, based on the RMT, a data-driven approach is developed for early anomaly detection and {localization} in distribution network. It leverages the similarities of the data collected from multiple measurement devices, and reveals the anomaly by tracking the variation of the data correlations. The primary contributions of this paper are shown as follows:
1) The approach is mainly data-driven and it only requires the simple topology information of feeder lines in the distribution network.
2) The approach merges anomaly detection and {localization} functionalities through analyzing the extreme eigenvalues (outliers) and the corresponding eigenvectors from the data.
3) The approach is sensitive to the variation of the data correlations, and it is capable of detecting and {localizing} the anomaly at an early stage.
4) It is experimentally validated that the approach is robust against random disturbance and measurement error.
5) An increasing data dimension algorithm is designed, which makes it more accurate for analyzing the low observability feeder lines in the distribution network.

{The other sections of this paper are arranged as follows.} {Section \ref{section: theory} presents the mathematical foundations of RMT for anomaly detection and {localization}, in which anomaly indicators are designed and analyzed.} In section \ref{section: application}, spatio-temporal matrices are formulated by leveraging the measurement data in distribution network and detailed steps of the early anomaly detection and {localization} approach are presented. {Meanwhile, an increasing data dimension algorithm is designed for analyzing the low observability feeders more accurately.} {In section \ref{section: case}, Both {MATPOWER simulation data and real SCADA data} are used to verify the feasibility of the developed approach . Conclusions and {future research directions} are illustrated in Section \ref{section: conclusion}.}

\section{Random Matrix Theory for Anomaly Detection and {Localization}}
\label{section: theory}
In practical world, massive amounts of data can be naturally represented by large random matrices \cite{Qiu2013Cognitive}. {In this section, we apply the RMT for anomaly detection and {localization} of high-dimensional data matrices. First, asymptotic theorem in the RMT is used to analyze the empirical spectral distribution (ESD) of high-dimensional `signal+noise' matrix, and linear statistics of the eigenvalues are introduced as a statistical index to track the data behavior. The details of the RMT for anomaly detection and {localization} are presented.}

\subsection{{Asymptotic Theorem for `Signal+Noise' Matrix}}
\label{subsection: statistical properties}
Marchenko-Pastur Law (M-P Law): {Assume ${\bf X}=\{{x}_{i,j}\}\in{{\mathbb C}^{p \times n}}$ being a random matrix with independent identically distributed (i.i.d.) entries satisfying: 1) the mean $\mu (x)=0$ and 2) the variance $\sigma ^2 (x)<\infty$. The covariance matrix of $\bf X$ is calculated as ${\bf \Sigma}=\frac{1}{n} {\bf X}{\bf X}^{H}$. According to the M-P law \cite{marvcenko1967distribution}, when $p,n \to\infty$ and $c=\frac{p}{n}\in (0,1]$,  the ESD of ${\bf \Sigma}$ fits to the theoretical limit with probability density function (PDF)
\begin{equation}
\label{Eq:mp-law}
\begin{aligned}
{f_{MP}}(\lambda) = \left\{ \begin{array}{l}
\frac{1}{{2\pi c{\sigma ^2}}\lambda}\sqrt {(b - \lambda)(\lambda - a)} {\rm{,}} \quad a \le\lambda \le b\\
0, \qquad  \qquad  \qquad  \qquad  \qquad   {\lambda < a \; \rm{or}\; \lambda > b}
\end{array} \right.
\end{aligned},
\end{equation}
where $a={\sigma ^2}{(1-\sqrt{c})}^2$, $b={\sigma ^2}{(1+\sqrt{c})}^2$.}

We apply the M-P law for a high-dimensional data matrix ${\bf X'}\in {\mathbb{R}}^{p\times n}$. In {steady} state, $\bf X'$ is considered to be a random matrix and the ESD of ${\bf \Sigma}'=\frac{1}{n} {\bf X'}{\bf X'}^{T}$ converges to the limiting spectral density ${f_{MP}}(x')$, {as} is shown in Figure \ref{fig:mplaw_mal}(a). The bars in blue color represent the eigenvalue distributions of ${\bf \Sigma}'$ and the M-P law is plotted in the red curve. However, what will happen in {unsteady} state? Here, ``{unsteady}'' means signals occur in ${\bf X'}$ and the correlations among the entries $x'_{ij}$ have been changed. Then ${\bf X'}$ is considered of the type ${\bf X}+{\bf P}$, where $\bf X$ is a random matrix which represents random noise or fluctuations, and $\bf P$ is a low-rank matrix which represents anomaly signals. Figure \ref{fig:mplaw_mal}(b) shows the ESD of ${\bf\Sigma}'$ does not converge to the M-P law. It can be observed that the outliers caused by anomaly signals are out of the range $[a,b]$ (i.e., $[0.034,3.301]$).
\begin{figure}[htb]
\centering
\begin{minipage}{4.1cm}
\centerline{
\includegraphics[width=1.8in]{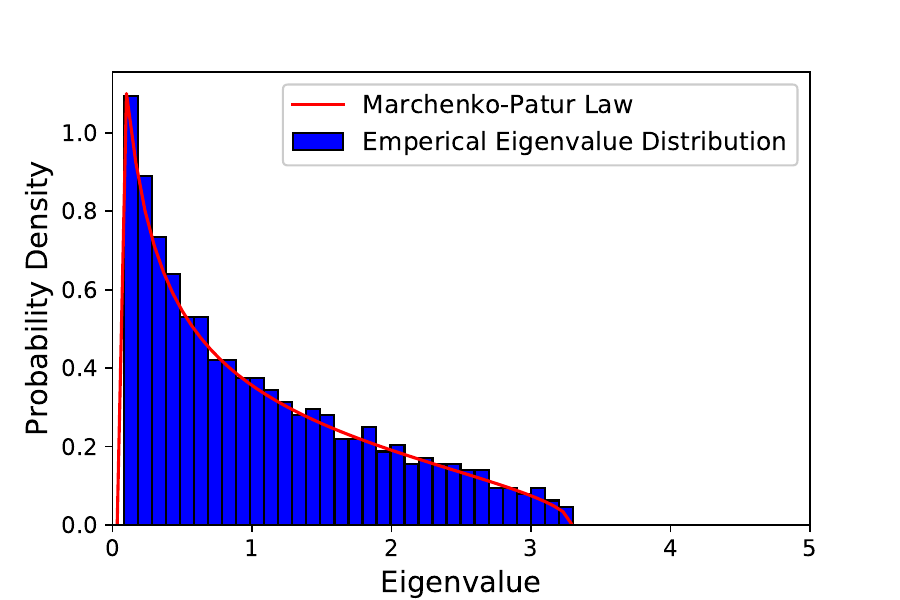}
}
\parbox{5cm}{\small \hspace{1.1cm}(a) Steady state }
\end{minipage}
\hspace{0.2cm}
\begin{minipage}{4.1cm}
\centerline{
\includegraphics[width=1.8in]{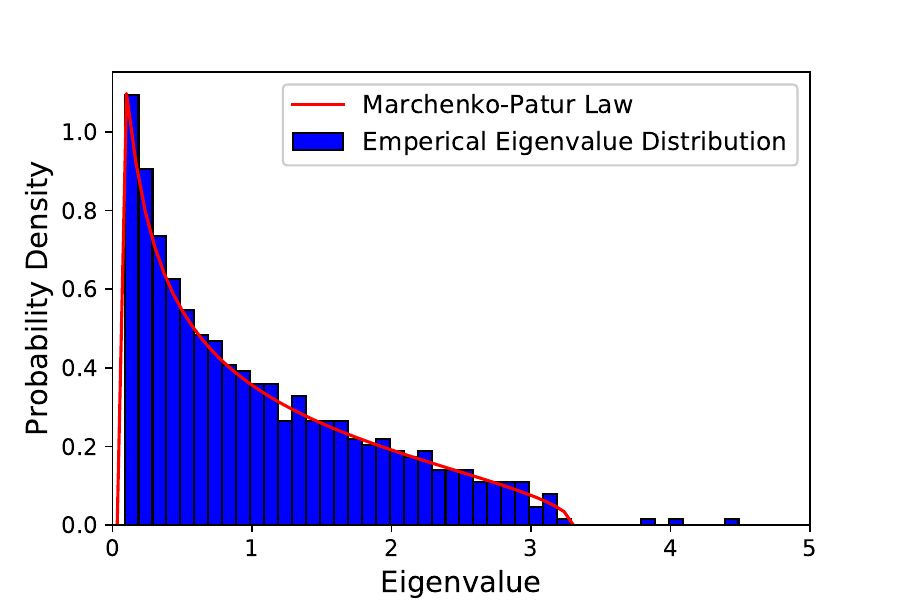}
}
\parbox{5cm}{\small \hspace{1.1cm}(b) Unsteady state }
\end{minipage}
\caption{The empirical spectral density of ${\bf\Sigma}'$ and its comparison with the theoretical limit corresponding to steady and unsteady system states, respectively. (a) $\bf X'$ is a $640\times 960$ random gaussian matrix. (b) ${\bf X'}={\bf X}+{\bf P}$, where $\bf X$ is a $640\times 960$ random gaussian matrix, and $\bf P$ is a low-rank signal matrix.}
\label{fig:mplaw_mal}
\end{figure}

Based on the analysis above, it can be concluded that the ESDs are different for a high-dimensional random matrix with or without anomaly signals, which inspires us to investigate the statistics regarding the empirical eigenvalues to indicate the data behavior. The linear eigenvalue statistics (LES) via test function $\phi$ are defined as
\begin{equation}
\label{Eq:les}
\begin{aligned}
  \mathcal{N}_\phi = \sum\limits_{i = 1}^n {\phi \left( {{\lambda _i}} \right)}
\end{aligned},
\end{equation}
where ${\lambda _i} (i=1,2,...,n)$ are the eigenvalues, and the test function ${\phi}$ is continuously smooth. {The test functions frequently used \cite{Qiu2013Cognitive} are listed as follows:}
\begin{itemize}
\item Chebyshev Polynomial (CP): $\phi (\lambda_i)={a_n}+{a_{n-1}}\lambda_i+\cdots+{a_0}{\lambda_i}^n$, where $a_k (k=1,2,\cdots,n)$ are real numbers;
\item Information Entropy (IE): $\phi (\lambda_i)=-{\lambda_i}ln{\lambda_i}$;
\item Likelihood Radio Function (LRF): $\phi (\lambda_i)=-1-ln{\lambda_i}+{\lambda_i}$;
\item Wasserstein Distance (WD): $\phi (\lambda_i)=1-2\sqrt{\lambda_i}+{\lambda_i}$.
\end{itemize}

The LES constructed via the listed test functions does not introduce any system error, which can be served as an statistical index to track the data behavior. It gives insight into the data behavior from a high-dimensional perspective, which makes it possible for detecting the latent anomalies in the data. Meanwhile, some statistical properties of the LES have been proved in theory \cite{shcherbina2011central}, such as satisfying the central limit theorem, with bounded variance, with a fast decay rate (i.e., in the order of $O(p^{-2})$) for the variance, etc.
\subsection{RMT for Anomaly Detection and {Localization}}
\label{subsection: detection location}
Assume there are $P-$dimensional measurement variables $(x_1,x_2,...,x_P)\in \mathbb{R}^P$ for each sampling time. At the sampling time $t_j$, the $P-$dimensional measurements can be formulated as a column vector ${\bf x}(t_j)=(x_1,x_2,...,x_P)^T$. For a series of time $N$, a data set $\bf D$ is formulated by arranging these vectors $\bf x$ in chronological order. Let $\bf X$ be a $p\times n$ { moving window} on $\bf D$, we can convert it into the standard form $\bf\hat X$ by
\begin{equation}
\label{Eq:standardize}
\begin{aligned}
  {\hat x_{ij}} = \left( {{x_{ij}} - \mu \left( {{{\bf x}_i}} \right)} \right) \times \frac{{\sigma \left( {{{\hat {\bf x}}_i}} \right)}}{{\sigma \left( {{{\bf x}_i}} \right)}} + \mu \left( {{{\hat {\bf x}}_i}} \right)
\end{aligned},
\end{equation}
where ${\bf x}_i=(x_{i1},x_{i2},...,x_{in})$, $\mu ({\bf\hat x}_i)=0$, and $\sigma ({\bf\hat x}_i)=1$ $(i=1,2,...,p;j=1,2,...,n)$. The covariance matrix of $\bf\hat X$ is calculated as ${\bf\Sigma}  = \frac{1}{n}{\bf{\hat X}}{{\bf{\hat X}}^T}$. Then the empirical eigenvalues $\lambda _{\bf\Sigma}$ and eigenvectors ${\bf v} _{\bf\Sigma}$ of ${\bf\Sigma}$ can be obtained.

The linear statistics of $\lambda _{\bf\Sigma}$ is calculated through equation (\ref{Eq:les}), which is served as the anomaly detection indicator in the developed approach. Furthermore, the anomaly is located based on the calculated $\lambda _{\bf\Sigma}$ and ${\bf v} _{\bf\Sigma}$. {According to the definitions on matrix eigenvalue and eigenvector}, it can be obtained}
\begin{equation}
\label{Eq:eigenvalue_vector}
\begin{aligned}
  {\bf{\Sigma }}{{\bf{v}}_{{\bf{\Sigma }},k}} = {\lambda _{{\bf{\Sigma }},k}}{{\bf{v}}_{{\bf{\Sigma }},k}}
\end{aligned}.
\end{equation}
The derivation of equation (\ref{Eq:eigenvalue_vector}) regarding the elements $\varepsilon_{ij}$ is
\begin{equation}
\label{Eq:derivative}
\begin{aligned}
  \frac{{d{\bf{\Sigma }}}}{{d{\varepsilon _{ij}}}}{{\bf{v}}_{{\bf{\Sigma }},k}} + {\bf{\Sigma }}\frac{{d{{\bf{v}}_{{\bf{\Sigma }},k}}}}{{d{\varepsilon _{ij}}}} = \frac{{d{\lambda _{{\bf{\Sigma }},k}}}}{{d{\varepsilon _{ij}}}}{{\bf{v}}_{{\bf{\Sigma }},k}} + {\lambda _{{\bf{\Sigma }},k}}\frac{{d{{\bf{v}}_{{\bf{\Sigma }},k}}}}{{d{\varepsilon _{ij}}}}
\end{aligned}.
\end{equation}
Since $\bf\Sigma$ is real and symmetric, and there exist ${{\bf v}_{\bf{\Sigma},k}}^{T}{{\bf v}_{\bf{\Sigma},k}}=1$. Left multiply ${{\bf v}_{{\bf\Sigma},k}}^{T}$ for equation (\ref{Eq:derivative}), we can obtain
\begin{equation}
\label{Eq:left_multiply}
\begin{aligned}
  \frac{{d{\lambda _{{\bf{\Sigma }},k}}}}{{d{\varepsilon _{ij}}}} = {{\bf{v}}_{{\bf{\Sigma }},k}}^T\frac{{d{\bf{\Sigma }}}}{{d{\varepsilon _{ij}}}}{{\bf{v}}_{{\bf{\Sigma }},k}}
\end{aligned},
\end{equation}
{where 
\begin{equation}
\label{Eq:differential}
\begin{aligned}
\frac{d{\bf\Sigma}}{d{\varepsilon_{ij}}} = \left\{ \begin{array}{l}
1 {\rm{,}} \qquad \varepsilon = {\varepsilon_{ij}} \\
0 {\rm{,}} \qquad \varepsilon \ne {\varepsilon_{ij}}
\end{array} \right.
\end{aligned}.
\end{equation}}
Thus we can simplify equation (\ref{Eq:left_multiply}) as
\begin{equation}
\label{Eq:simplify}
\begin{aligned}
  \frac{{d{\lambda _{{\bf{\Sigma }},k}}}}{{d{\varepsilon _{ij}}}} = v_{{\bf{\Sigma }},k}^{(i)}v_{{\bf{\Sigma }},k}^{(j)}
\end{aligned}.
\end{equation}
The contribution rate of the entries in the $i$th row of $\bf\Sigma$ to $\lambda_{{\bf\Sigma},k}$ can be calculated as
\begin{equation}
\label{Eq:contribution}
\begin{aligned}
  \sum\limits_{j = 1}^p {(\frac{{d{\lambda _{{\bf{\Sigma }},k}}}}{{d{\varepsilon _{ij}}}}} {)^2} = {(v_{{\bf{\Sigma }},k}^{(i)})^2}\sum\limits_{j = 1}^p {{{(v_{{\bf{\Sigma }},k}^{(j)})}^2}} = {(v_{{\bf{\Sigma }},k}^{(i)})^2}
\end{aligned}.
\end{equation}

From equation (\ref{Eq:contribution}), it can be concluded that the $i$th entry of ${\bf v}_{{\bf{\Sigma}},k}$ can be used to measure the ``contribution'' of the $i$th row of $\bf\Sigma$ to $\lambda_{{\bf\Sigma},k}$. Based on the analysis of the M-P law for high-dimensional `signal+noise' matrix in Section \ref{subsection: statistical properties}, it can be observed that outliers (i.e., $\lambda > b$) occur when a system operates in {unsteady} state. {Thus we can realize anomaly localization by analyzing the eigenvectors corresponding to the outliers.} The anomaly {localization} indicator can be designed as
\begin{equation}
\label{Eq:location_indicator}
\begin{aligned}
  {\eta _i} = \frac{{\sum\limits_{{\lambda _{{\bf{\Sigma }},k}} \in \{\lambda > b \}} {{\lambda _{{\bf{\Sigma }},k}}{{(v_{{\bf{\Sigma }},k}^{(i)})}^2}} }}{{\sum\limits {{\lambda _{{\bf{\Sigma }},k}}} }}
\end{aligned},
\end{equation}
where $\eta _i\in [0,1)$. The indicator $\eta_i$ measures the scale of the $i$th row's ``contribution'' to the anomaly.

We first standardize $\bm\eta\; {(\eta_i\in{\bm\eta})}$ by
\begin{equation}
\label{Eq:location_indicator_standard}
\begin{aligned}
  \hat{\bm\eta}=\frac{\bm\eta-\mu({\bm\eta})}{\sigma({\bm\eta})}
\end{aligned},
\end{equation}
where $\mu({\bm\eta})$ is the mean and $\sigma({\bm\eta})$ represents the standard deviation of $\bm\eta$, and $\hat{\bm\eta}({\hat{\eta_i}}\in{\hat{\bm\eta}})$ is the standardized $\bm\eta$. Considering the sample size $p$ of $\hat{\bm\eta}$ is sometimes small, here, $\hat{\bm\eta}$ is assumed to be approximately a t distribution with $p-1$ freedom degree. According to the central limit theorem, the confidence level $1-\alpha$ for the population mean $\mu$ of $\hat{\bm\eta}$ is defined as
\begin{equation}
\label{Eq:confidence_level}
\begin{aligned}
  1-\alpha = P\{{\mu(\hat{\bm\eta})-t_{\frac{\alpha}{2}}{\frac{\sigma(\hat{\bm\eta})}{\sqrt{p}}}}<\mu<{\mu(\hat{\bm\eta})+t_{\frac{\alpha}{2}}{\frac{\sigma(\hat{\bm\eta})}{\sqrt{p}}}}\}
\end{aligned},
\end{equation}
where $\mu(\hat{\bm\eta})$is the sample mean with $\mu(\hat{\bm\eta})=0$ and $\sigma(\hat{\bm\eta})$ is the standard deviation of $\hat{\bm\eta}$ with $\sigma(\hat{\bm\eta})=1$, $t_{\frac{\alpha}{2}}$ is the upper $\frac{\alpha}{2}$ critical value for the t distribution, and $P\{\cdot\}$ is the probability operator. For a given $\hat{\eta_i}$, the corresponding confidence level $1-\alpha$ can be obtained by the t distribution table. {For example, let $\hat{\eta_i}=2.064$ and $p=25$, then the value of $1-\alpha$ is calculated to be $95\%$.} Thus, the anomaly can be {localized} through comparing $1-\alpha$ with the pre-defined threshold $(1-\alpha)_{th}$.

In real-time analysis, we can move a window on the formulated data set $\bf D$ continuously to track the data behavior. {Take the current sampling time $t_j$ as an example, the generated data matrix ${\bf X}(t_j)$ is written as}
\begin{equation}
\label{Eq:matrix_formulate}
\begin{aligned}
  {\bf{X}}(t_j) = \left( {{\bf{x}}(t_{j - n + 1}),{\bf{x}}(t_{j - n + 2}), \cdots ,{\bf{x}}(t_j)} \right)
\end{aligned},
\end{equation}
where ${\bf x}(t_k)={({x_1,x_2,\cdot\cdot\cdot,{ x_p}})}^T$ for $t_{j-n+1}\le t_k \le t_j$ represents the measurement data at the sampling time $t_k$. Thus, the indicator $\mathcal{N}_{\phi}(t_j)$ and ${\bm\eta}({t_j})$ can be calculated for the current sampling time $t_j$.

\section{Early Anomaly Detection and {Localization} in Distribution Network}
\label{section: application}
In this section, by leveraging {the measurement data collected from the SCADA system installed in distribution network}, a RMT-based early anomaly detection and {localization} approach is developed. First, the measurement data from multiple monitoring devices for each feeder line is formulated as a spatio-temporal data set. Then, detailed steps of the developed approach are presented and advantages of the approach are remarked. Last, an increasing data dimension algorithm is designed for the low observability feeders being more accurately analyzed.
\subsection{Formulation of Measurement Data as Spatio-Temporal Matrices}
\label{subsection: data_form}
\begin{figure}[htb]
\centerline{
\includegraphics[width=2.5in]{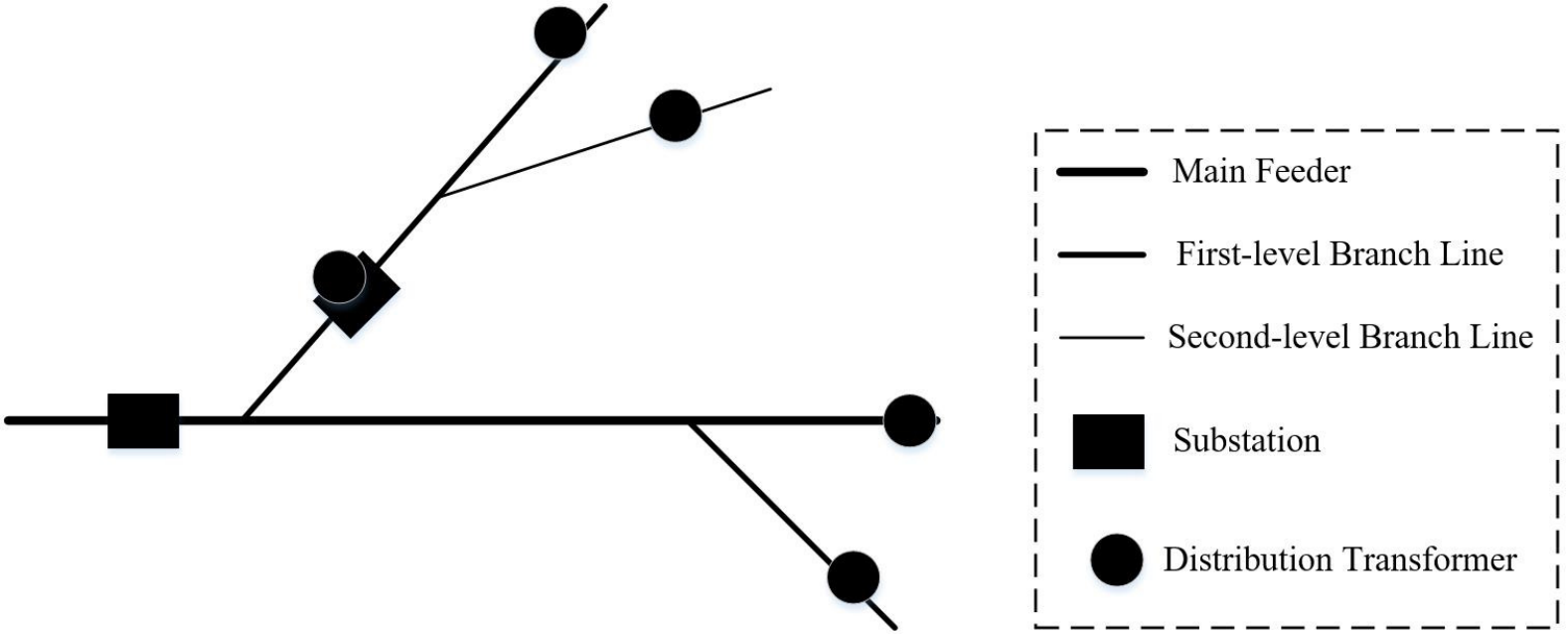}
}
\caption{Circuitry topology diagram of partial distribution network.}
\label{fig:circuitry_topology}
\end{figure}
Figure \ref{fig:circuitry_topology} illustrates circuitry topology diagram of partial distribution network, in which a feeder line consists of different level of branch lines and substations with distribution transformers. Multiple online monitoring devices are installed at different physical locations of the feeder, through which we can obtain many types of measurement variables, such as three-phase voltage ($u_a,u_b,u_c$), three-phase current ($i_a,i_b,i_c$), active load ($l$), etc. Here, $7$ measurements ($u_a,u_b,u_c,i_a,i_b,i_c,l$) at the sampling time $t_j$ are chosen as the elements to formulate a data vector ${{\bf{d}}(t_j)} = {\left[ {{{u}}_{aj}^{\left( 1 \right)},{{u}}_{bj}^{\left( 1 \right)},{{u}}_{cj}^{\left( 1 \right)},{{i}}_{aj}^{\left( 1 \right)},{{i}}_{bj}^{\left( 1 \right)},{{i}}_{cj}^{\left( 1 \right)},{{l}_{j}}^{(1)}, \cdots ,{{l}}_{j}^{\left( m \right)}} \right]^T}$, where $m$ denotes the number of monitoring devices and {${{{u}}_{aj}^{\left( k \right)},{{u}}_{bj}^{\left( k \right)},{{u}}_{cj}^{\left( k \right)},{{i}}_{aj}^{\left( k \right)},{{i}}_{bj}^{\left( k \right)},{{i}}_{cj}^{\left( k \right)}}(k=1,\cdots,m)$ are the root mean square (RMS) values}. Assume $P=7m$, for a series of time $N$, we can obtain the data set ${\bf D}=[{\bf d}(t_1),{\bf d}(t_2),\cdots,{\bf d}(t_N)]\in{\mathbb{R}^{P\times N}}$. {It is noted that, by stacking the measurements together, the formulated spatio-temporal data set contains rich information on the feeder operating states.}

\subsection{Early Anomaly Detection and {Localization}}
\label{subsection: approach_step}
{Based on the work above, we develop a new approach for early anomaly detection and {localization} in distribution network.} The specific steps are given in {Table \ref{Tab: algorithm}}.
\begin{table}[htbp]
\caption{}
\label{Tab: algorithm}
\centering

\begin{tabular}{p{8.4cm}}   
\toprule[1.0pt]
\textbf {Steps of the RMT for early anomaly Detection and {Localization} in Distribution Network}\\
\hline
1: {A spatio-temporal data set ${\bf D}\in{\mathbb{R}^{P\times N}}$ is formulated for every feeder}\\
\quad by arranging $P$ measurements in a series of time $N$.  \\
2: At the sampling time $t_j$: \\
  \quad 2a) Form the data matrix ${\bf X}(t_j)$ by using a $p\times n$ ($p=P,n<N$) \\
  \quad\quad\; { window} on $\bf D$; \\
  \quad 2b) Convert ${\bf X}(t_j)$ into the standard form matrix ${\hat{\bf X}}(t_j)$ through \\
  \quad\quad\; equation (\ref{Eq:standardize}); \\
  \quad 2c) Calculate the sample covariance matrix of ${\hat{\bf X}}(t_j)$, i.e., ${\bf\Sigma}(t_j)$;  \\
  \quad 2d) Obtain the eigenvalues $\lambda_{\bf\Sigma}(t_j)$ and eigenvectors ${\bf v}_{\bf\Sigma}(t_j)$, \\
  \quad\quad\; and compare the ESD with the theoretical limits; \\
  \quad 2e) Calculate the linear statistics of $\lambda_{\bf\Sigma}(t_j)$ through equation (\ref{Eq:les}), \\
  \quad\quad\; i.e., $\mathcal{N}_\phi(t_j)$; \\
  \quad 2f) Calculate the {localization} indicator ${\bm\eta}(t_j)$ by using equation (\ref{Eq:location_indicator}). \\
3: Plot the $\mathcal{N}_{\phi}-t$ curve for every feeder during a period of time $N$. \\
4: Plot the 3D graph of $\bm\eta$ regarding the $P$ measurements and time $N$, \\
\quad and calculate the value of $1-\alpha$ for every point of the graph to {localize} \\
\quad the anomaly indexes. \\
\hline
\end{tabular}
\end{table}

{The early anomaly detection and {localization} approach is driven by the {measurement data from SCADA system installed in distribution network}. It is sensitive to the variation of the data correlations, and it is capable of detecting and {localizing} the anomaly at an early stage.} The steps above involve no mechanism models, thus avoiding the errors brought by assumptions and simplifications. The approach merges anomaly detection and {localization} functionalities, and experimentally robust to random disturbance and measurement error of the data. {The computational time of the approach is mainly determined by the calculation of covariance matrix and eigenvalue in Step 2c and 2d, complexity of which approximates O($p^2n^2$). In practice, the covariance matrix and eigenvalue calculation can be implemented by using $numpy.cov()$ and $numpy.linalg.eig()$ functions in Python, which has an extremely fast computing rate}. Therefore, the approach is practical for both online and offline analysis.

\subsection{Discussion}
\label{subsection: discussion}
We may notice that the M-P law holds when the data dimensions are infinite or large. However, in the application of early anomaly detection and {localization} in distribution network, there exist some low observability feeders. Dimensions of the formulated data matrices corresponding to those feeders are often moderate, such as tens or less. In \cite{Qiu2017} and \cite{ambainis2012random}, a natural way of increasing dimensions of data vectors based on tensor product is introduced. On this basis, here, an algorithm to increase the dimensions of data matrices is designed. The algorithm allows for the analysis of high-dimensional data matrices and yields smaller variance for the related functionals.

Assume ${\bf X}=[{\bf x}_1,{\bf x}_2,\cdots,{\bf x}_t]\in\mathbb{C}^{p\times t}$ be a random matrix with i.i.d. entries, $p=kn$. For $k,t,n\in\mathbb{N}$, we construct a new random vector by using the tensor product of the column vectors of $\bf X$ in the form
\begin{equation}
\label{Eq:Kron}
\begin{aligned}
  {{\bf\tilde{x}}_j} = {\bf{x}}_j^{(1)} \otimes  \cdots  \otimes {\bf{x}}_j^{(k)} \in {\left( {{\mathbb{C}^n}} \right)^{ \otimes k}}
\end{aligned},
\end{equation}
where ${\bf x}_j^{(l)}(j=1,2,\cdots,n; l=1,2,\cdots,k)$ are i.i.d. copies of a normalized isotropic random vector ${\bf x}_j^{(1)}=( x_{1j},\cdots,x_{nj})\in {\mathbb{C}^n}$, and `$\otimes$' denotes the tensor product operation. {Here, `isotropic' indicates the vectors have the same mean and variance}. The new random vector ${\bf\tilde{x}}_j$ lies in the $n^k$ dimensional normed space. Thus, the corresponding dimension increased random matrix ${\bf\tilde X}=[{\bf\tilde x}_1,{\bf\tilde x}_2,\cdots,{\bf\tilde x}_t]\in\mathbb{C}^{n^k\times t}$ is obtained. The time and space complexity of the algorithm are O($kt$) and O($n^k$).

Now consider $n^k\times n^k$ random matrices of the form
\begin{equation}
\label{Eq:Kron_covariance}
\begin{aligned}
  {{\bf{\mathcal{M}}}_{n,t,k}}\left( {\bf{x}} \right) = \sum\limits_{\alpha = 1}^t {{\tau _\alpha}{{{\bf{\tilde x}}}_\alpha}{\bf{\tilde x}}_\alpha^H}
\end{aligned},
\end{equation}
where $\tau_\alpha(\alpha=1,2,\cdots,t)$ are real numbers. The asymptotic behavior of ${{\bf{\mathcal{M}}}_{n,t,k}}({\bf{x}})$ has been well studied in \cite{lytova2017central}. For every fixed $k\ge 1$, as $t\to\infty, n\to\infty,$ but $\frac{t}{n^k} \to c\in (0,\infty)$, the ESD of ${{\bf{\mathcal{M}}}_{n,t,k}}({\bf{x}})$ converges to a non-random measure.

\begin{figure}[htb]
\centering
\begin{minipage}{4.1cm}
\centerline{
\includegraphics[width=1.90in]{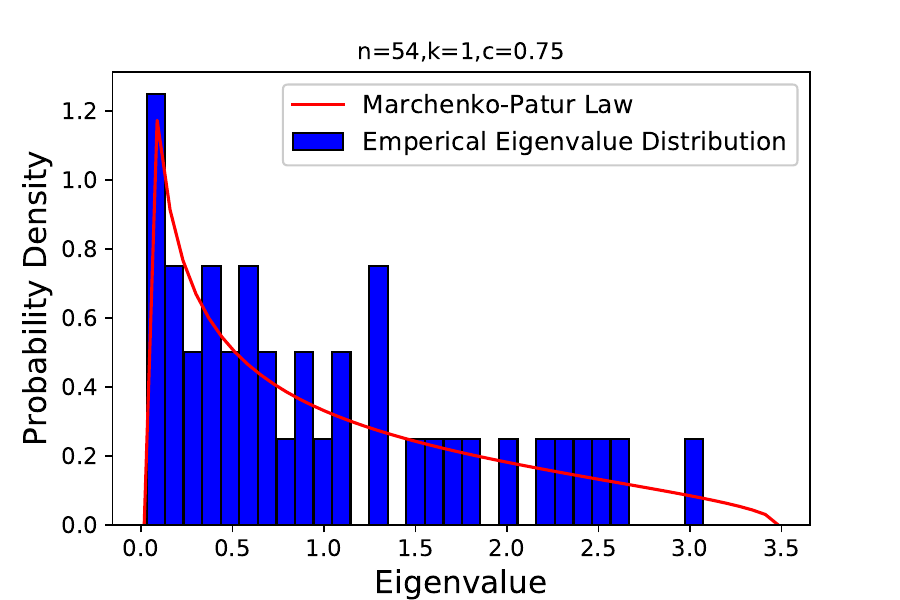}
}
\parbox{5cm}{\small \hspace{1.2cm}(a) Dimension: 54 }
\end{minipage}
\hspace{0.2cm}
\begin{minipage}{4.1cm}
\centerline{
\includegraphics[width=1.90in]{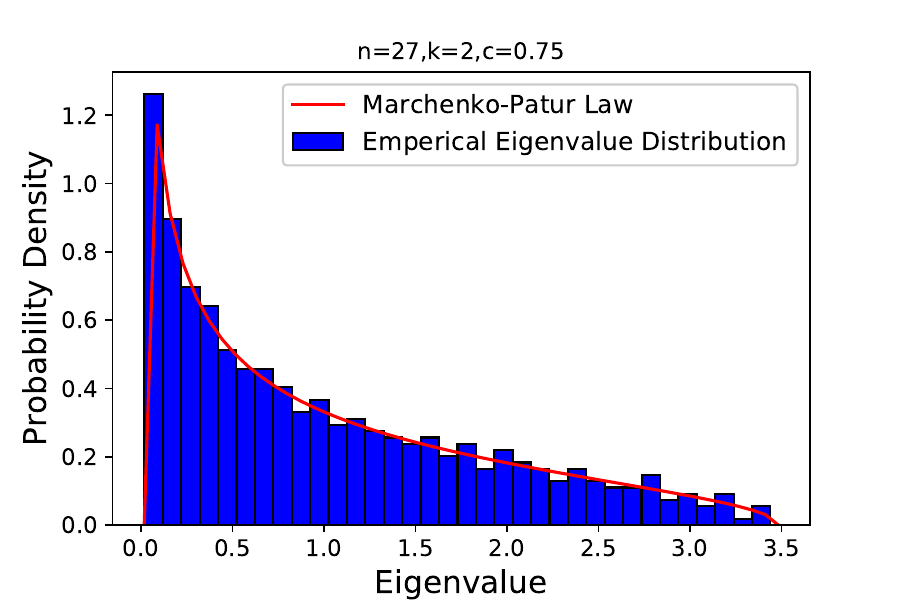}
}
\parbox{5cm}{\small \hspace{1.2cm}(b) Dimension: 729 }
\end{minipage}
\hspace{0.2cm}

\caption{The ESD of ${{\bf{\mathcal{M}}}_{n,t,k}}({\bf{x}})$ and its comparison with the M-P law. The dimension of $\bf X$ is increased from $27\times 2=54$ to $27^2=729$ through a tensor product of its column vectors, the radio $c$ is $0.75$, and $\tau _\alpha$ is $1$.}
\label{fig:mplaw_tensor}
\end{figure}
The ESD of the original covariance random matrix and the tensor product version and their comparisons with the theoretical limits are plotted in Figure \ref{fig:mplaw_tensor}(a) and Figure \ref{fig:mplaw_tensor}(b), respectively. It can be observed that the ESD of the original covariance random matrix does not fit to the M-P law for the reason of low dimensions. In contrast, the ESD of the tensor product version of covariance random matrix converges almost surely to the theoretical limits. The increasing data dimension algorithm makes it more accurate for analyzing the formulated data matrices from low observability feeders in the distribution network.

{The other issue we want to discuss about the developed approach is that the M-P law holds for the i.i.d. data matrix, which is not true for the measurements collected from a real-world power system. In practice, the collected measurements are correlated whether the system is in steady or unsteady state, and the M-P law does not hold true for them. The differences lie that, the correlations of the measurements corresponding to steady system state are much weaker than that in unsteady state. Thus, the spectrum from the collected measurement data are different when the system operates in steady or unsteady state and the M-P law can be used for depicting the differences, which guarantees the feasibility of the approach for the real data analysis in theory.}

\section{Case Studies}
\label{section: case}
In this section, the effectiveness of the developed approach is validated by using both the simulation data generated from standard IEEE test systems \cite{5491276} and the real {measurement data collected from the SCADA system installed in a distribution network.} Six cases in different scenarios are designed: 1) In the first four cases, by using the synthetic data, we test the effectiveness of our approach with different test functions, the designed increasing data dimension algorithm, the anomaly {localization} function and the advantages of the developed approach. 2) The last two cases, leveraging the real measurement data, validate the effectiveness of our approach for analyzing both high and low observability feeders in the distribution network.

\subsection{Case Study with Simulation Data}
\label{subsection: case_synthetic}
{The simulation data was generated from IEEE 57-bus and 33-bus test systems \cite{5491276}. The IEEE 33-bus test system is a standard distribution test network and the IEEE 57-bus test system is considered as a distribution network connected to generators. See case57.m and case33.m in Matpower \cite{zimmerman2016matpower} for details}. For the generated data $\bf D$, a little noise $\bf E$ was introduced to play the role of random disturbance and measurement error, i.e., ${\bf D}={\bf D}+\gamma{\bf E}$. The scale of the added noise is $\gamma=\sqrt{\frac{Tr({\bf D}{\bf D}^H)}{Tr({\bf E}{\bf E}^H)\times{\tau_{SNR}}}}$, where $Tr()$ represents the trace function and $\tau_{SNR}$ is the signal-to-noise ratio. In case 1 and case 2, the white noise was introduced, i.e., $E\sim N(0,1)$; in case 3 and case 4, the colored noise was introduced, i.e., $E_{it}=0.5*E_{i,t-1}+\varepsilon_{it}$, where $\varepsilon_{it}\sim N(0,1-0.5^2)$ so that the variance of $E_t$ is $1$.

\begin{figure}[!t]
\centerline{
\includegraphics[width=3.0in]{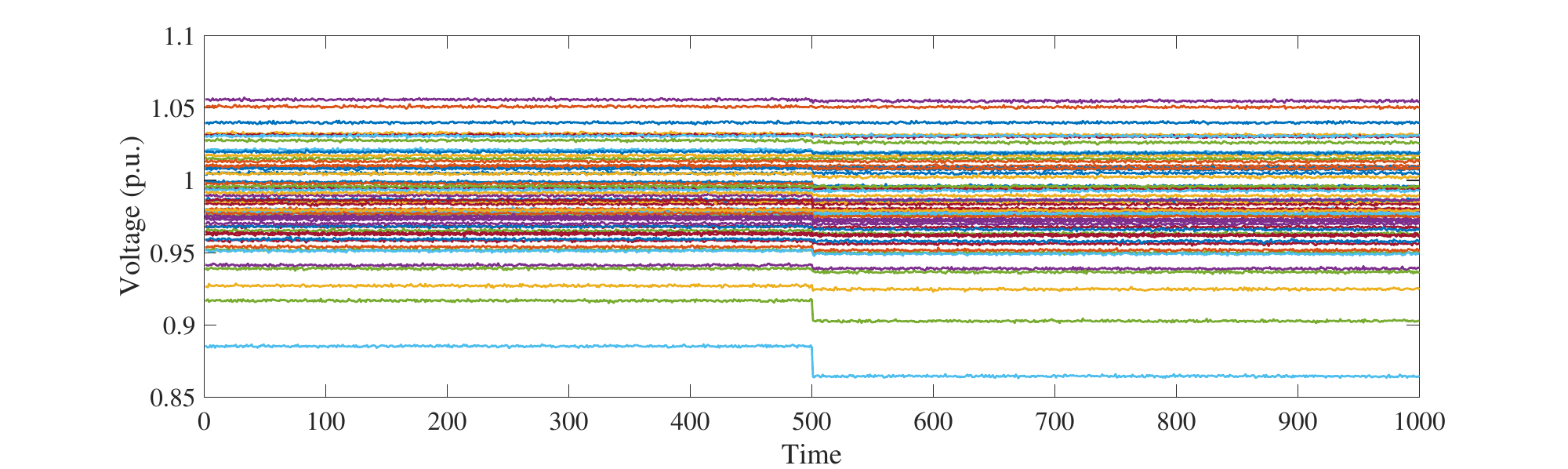}
}
\caption{{The generated voltage measurement curves in Case 1.}}
\label{fig:case1_data_org}
\end{figure}

1) Case Study on Different Test Functions: In this case, the IEEE 57-bus test system was used to {produce the simulation data}. For testing the effectiveness of the developed approach with different test functions in equation (\ref{Eq:les}), an anomaly signal was set by  changing the active load at bus $20$ suddenly, as shown in Table \ref{Tab: Case1}. The generated data consisted of $57$ voltage measurements for a series of $1000$ sampling times. The {voltage measurement curves} were plotted in Figure \ref{fig:case1_data_org}. In the experiment, the moving window's size was $57\times 200$ and $\tau_{SNR}$ was set to be $500$.
\begin{table}[!t]
\caption{{The Anomaly Signal Set at Bus $20$ in Case 1.}}
\label{Tab: Case1}
\centering
\begin{tabular}{clc}
\toprule[1.0pt]
\textbf{Bus} & \textbf{Sampling Time}& \textbf{Active Load(MW)} \\
\hline
\multirow{2}*{20} & $t_s=1\sim 500$ & 10 \\
~&$t_s=501\sim 1000$ & 12 \\
\hline
Others & $t_s=1\sim 1000$ & Unchanged \\
\hline
\end{tabular}
\end{table}
\begin{figure}[!t]
\centerline{
\includegraphics[width=2.5in]{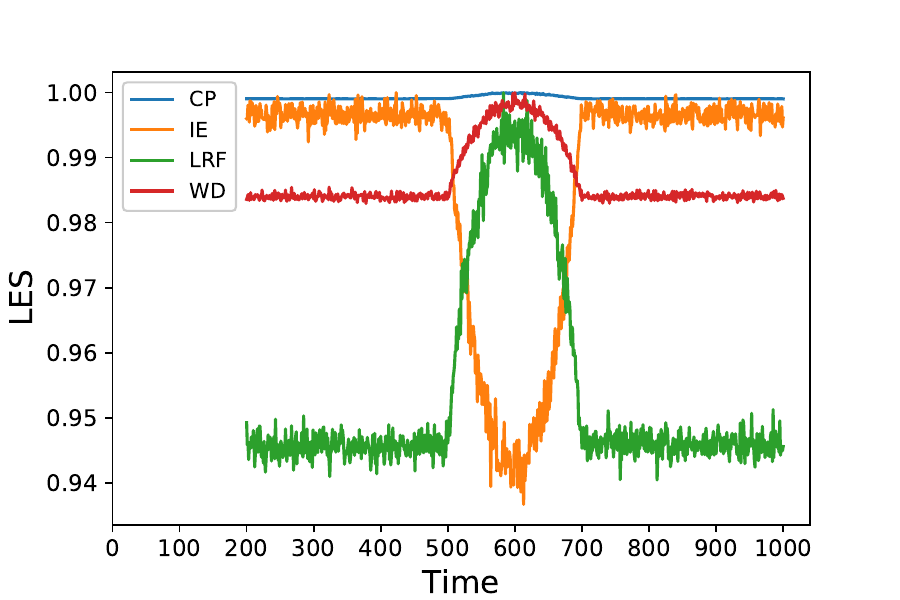}
}
\caption{{Effectiveness of the developed approach with different test functions.}}
\label{fig:test_function}
\end{figure}

\begin{figure}[!t]
\centering
\begin{minipage}{4.1cm}
\centerline{
\includegraphics[width=1.90in]{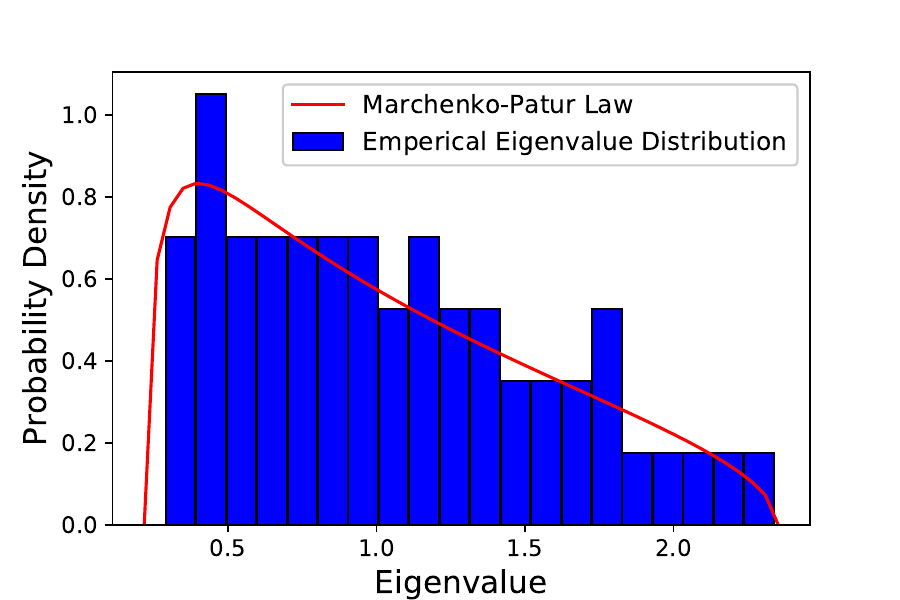}
}
\parbox{5cm}{\small \hspace{1.5cm}(a) $t_s=500$}
\end{minipage}
\hspace{0.2cm}
\begin{minipage}{4.1cm}
\centerline{
\includegraphics[width=1.90in]{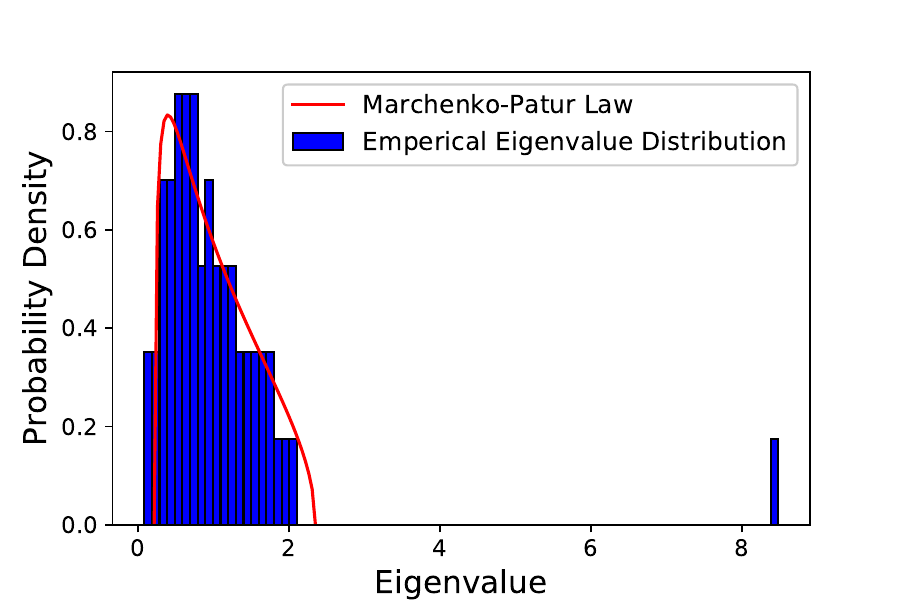}
}
\parbox{5cm}{\small \hspace{1.5cm}(b) $t_s=501$}
\end{minipage}
\caption{{The ESDs from the voltage measurements and their comparisons with the theoretical M-P law in Case 1.}}
\label{fig:case1_law}
\end{figure}

The anomaly detection results of the developed approach with different test functions are normalized into $[0,1]$, as shown in Figure \ref{fig:test_function}. It is noted that the $\mathcal{N}_\phi-t$ curve begins at $t_s=200$, because the moving data window consists of 199 historical samples and the current sample. From the $\mathcal{N}_\phi-t$ curves, it can be observed:

\uppercase\expandafter{\romannumeral1}. During $t_s=200\sim 500$, $\mathcal{N}_\phi$ computed through the developed approach with 4 different test functions remain nearly constant, which denotes the system is in {steady} state. As shown in Figure \ref{fig:case1_law}(a), the ESD converges almost surely to the theoretical M-P law.

\uppercase\expandafter{\romannumeral2}. From $t_s=501$, $\mathcal{N}_\phi$ begin to change dramaticlly, which denotes an anomaly occurs. Figure \ref{fig:case1_law}(b) shows that there exists one outlier, which coincides with the anomaly signal set during the data generation process. It is noted that, from $t_s=501\sim 700$, the $\mathcal{N}_\phi-t$ curves are almost U-shaped or inverted U-shaped, because {the data window is moved continuously and the duration of the anomaly signal on $\mathcal{N}_\phi$ is determined by the window width}. What's more, the $\mathcal{N}_\phi-t$ curve corresponding to test function $IE$ has the highest variance radio, which indicates the anomaly indicator via test function $IE$ is more sensitive to the abnormal data behavior. Hence $IE$ was chosen as the test function in subsequent cases.

\begin{figure}[!t]
\centerline{
\includegraphics[width=3.0in]{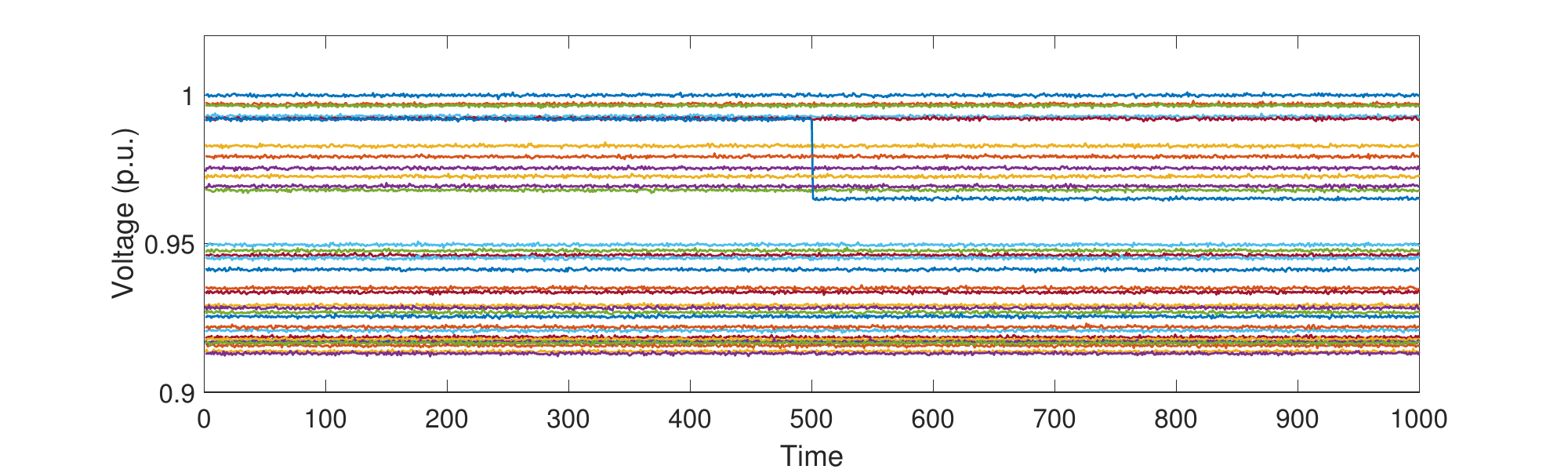}
}
\caption{{The generated voltage measurement curves in Case 2.}}
\label{fig:case2_data_org}
\end{figure}

2) Case Study on Increasing Data Dimension: In this case, the effectiveness of the designed increasing data dimension algorithm was tested. The IEEE 33-bus test system was used to {produce the simulation data}. During the simulation, an anomaly signal was set by increasing the impedance from bus $21$ to $22$ suddenly, as shown in Table \ref{Tab: Case2}. The generated simulation data consisted of {$33$} voltage measurements for a series of $1000$ sampling times. The {voltage measurement curves} were plotted in Figure \ref{fig:case2_data_org}. In the experiment, the moving window's size was $33\times 200$, the parameter $\tau_\alpha$ in equation (\ref{Eq:Kron_covariance}) was $1$, and the signal-to-noise ratio $\tau_{SNR}$ defined in Case 1 was set as $500$. By using the designed increasing data dimension algorithm, the dimension of each data window was increased from $33=16+17$ to $272=16\times 17$.
\begin{table}[!t]
\caption{{The Anomaly Signal Set From Bus $21$ to $22$ in Case 2.}}
\label{Tab: Case2}
\centering
\begin{tabular}{cclc}   
\toprule[1.0pt]
\textbf {fBus} & \textbf {tBus} & \textbf{Sampling Time}& \textbf{Impedance(p.u.)}\\
\hline
\multirow{2}*{21} & \multirow{2}*{22} & $t_s=1\sim 500$ & 0.5 \\
~ & ~ & $t_s=501\sim 1000$ & 20 \\
\hline
Others & Others & $t_s=1\sim 1000$ & Unchanged \\
\hline
\end{tabular}
\end{table}
\begin{figure}[!t]
\centering
\begin{minipage}{4.1cm}
\centerline{
\includegraphics[width=1.85in]{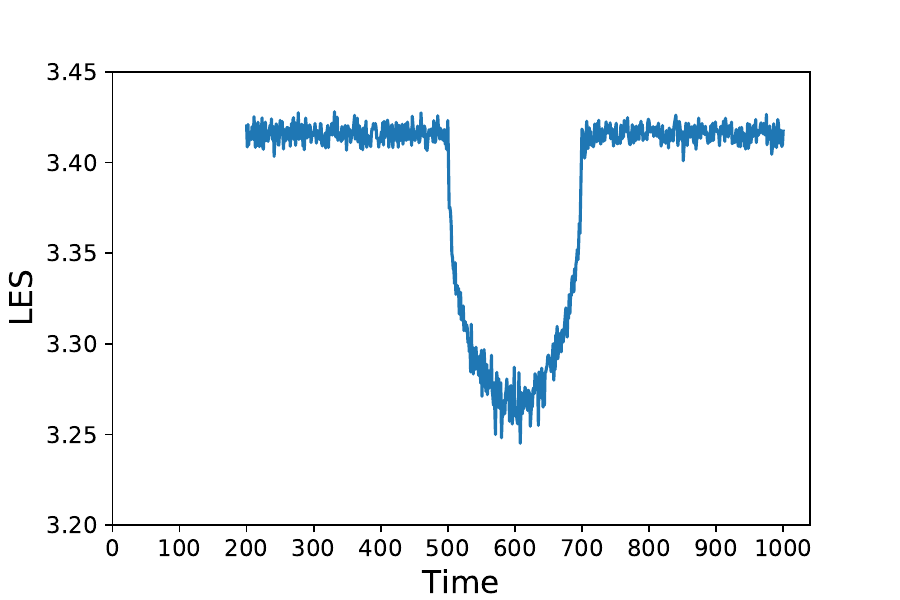}
}
\parbox{5cm}{\small \hspace{1.5cm}(a) Dimension:$33$}
\end{minipage}
\hspace{0.2cm}
\begin{minipage}{4.1cm}
\centerline{
\includegraphics[width=1.85in]{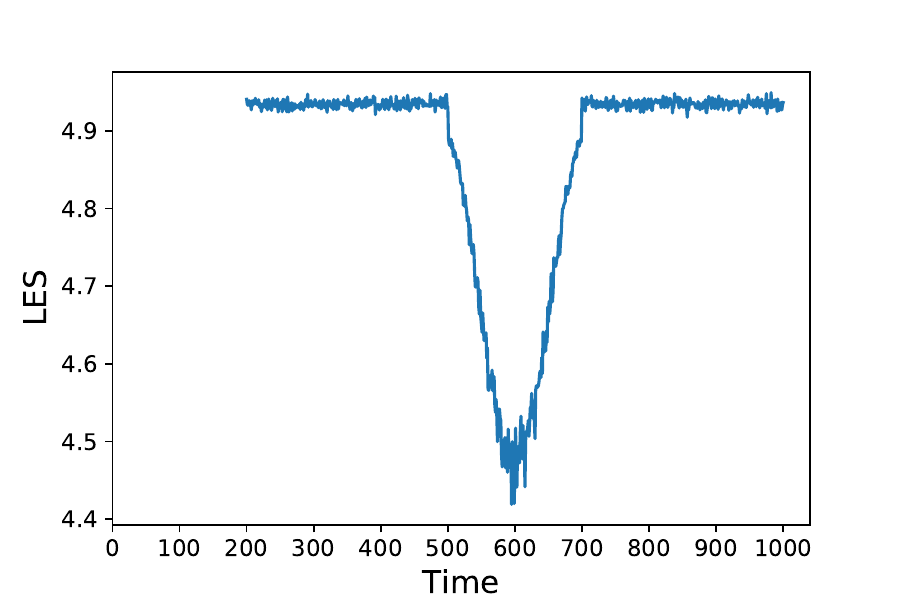}
}
\parbox{5cm}{\small \hspace{1.5cm}(b) Dimension:$272$}
\end{minipage}
\caption{{The anomaly detection results corresponding to different data dimensions. The dimension of the data was increased from $33$ to $272$.}}
\label{fig:case2_les}
\end{figure}

\begin{figure}[!t]
\centering
\begin{minipage}{4.1cm}
\centerline{
\includegraphics[width=1.90in]{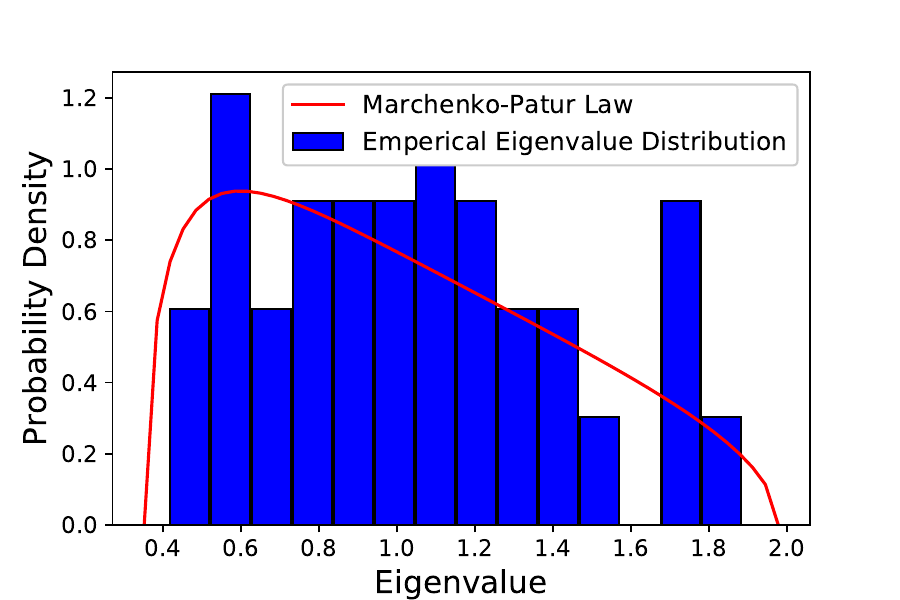}
}
\parbox{5cm}{\small \hspace{0.2cm}(a) $t_s=500$ (Dimension:33)}
\end{minipage}
\hspace{0.2cm}
\begin{minipage}{4.1cm}
\centerline{
\includegraphics[width=1.90in]{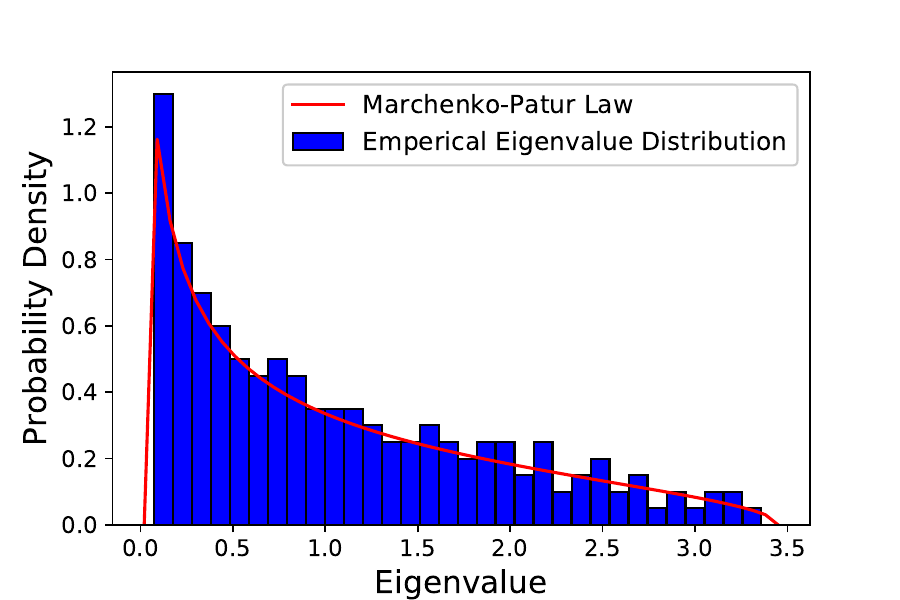}
}
\parbox{5cm}{\small \hspace{0.2cm}(b) $t_s=500$ (Dimension:272)}
\end{minipage}
\hspace{0.2cm}
\begin{minipage}{4.1cm}
\centerline{
\includegraphics[width=1.90in]{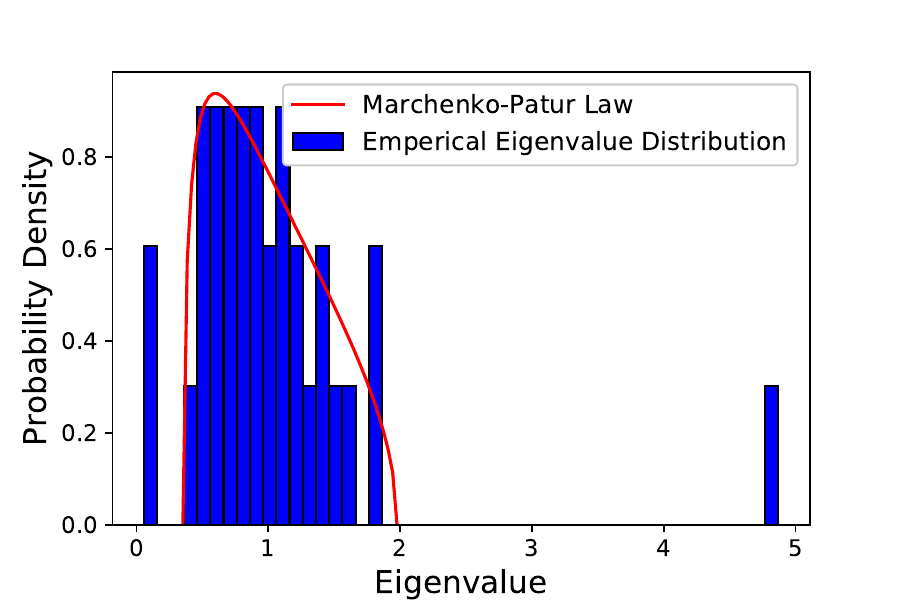}
}
\parbox{5cm}{\small \hspace{0.2cm}(c) $t_s=600$ (Dimension:33)}
\end{minipage}
\hspace{0.2cm}
\begin{minipage}{4.1cm}
\centerline{
\includegraphics[width=1.90in]{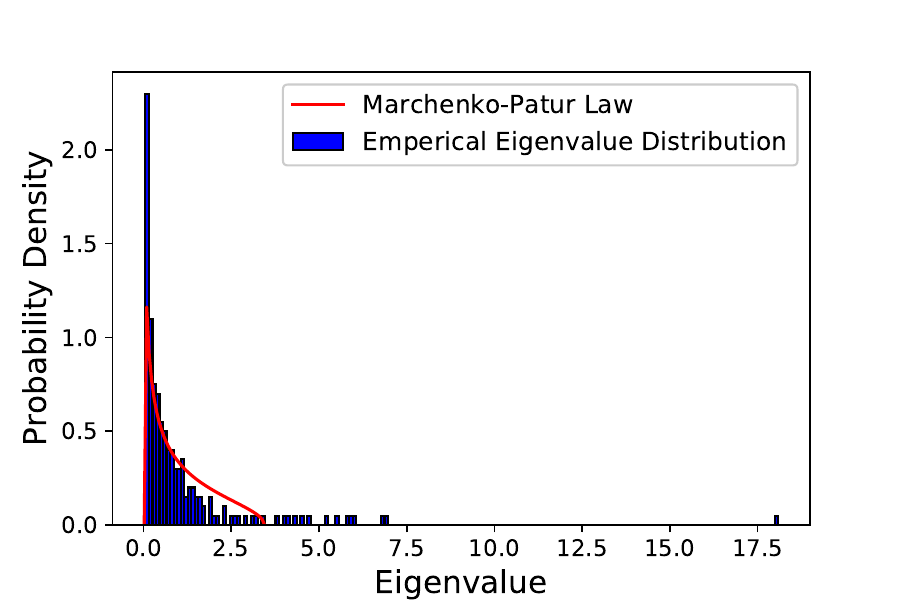}
}
\parbox{5cm}{\small \hspace{0.2cm}(d) $t_s=600$ (Dimension:272)}
\end{minipage}
\caption{{The ESDs from the voltage measurements and their comparisons with the theoretical M-P law in Case 2.}}
\label{fig:case2_law}
\end{figure}

Figure \ref{fig:case2_les}(a) and \ref{fig:case2_les}(b) show the anomaly detection results corresponding to different data dimensions. From the $\mathcal{N}_\phi-t$ curves, it can be obtained:

\uppercase\expandafter{\romannumeral1}. During $t_s=200\sim 500$, the values of $\mathcal{N}_\phi$ remain almost at $3.33$, $4.76$, respectively, which denotes the system is in {steady} state. The ESD does not fit to the theoretical M-P law in Figure \ref{fig:case2_law}(a) for the reason of low data dimension. In contrast, as shown in Figure \ref{fig:case2_law}(b), the ESD converges almost surely to the theoretical limit after the data's dimension is increased from $33$ to $225$.

\uppercase\expandafter{\romannumeral2}. From $t_s=501$, the $\mathcal{N}_\phi-t$ curves begin to change dramaticlly, which denotes an anomaly signal occurs and the system is in {unsteady} state. For example, during $t_s=501\sim 600$, the values of $\mathcal{N}_\phi$ reduce almost from $3.32$, $4.75$ to $3.10$, $2.80$, respectively. The ESD does not fit the theoretical M-P law for the reason of existing outliers, {as} shown in Figure \ref{fig:case2_law}(c) and \ref{fig:case2_law}(d). The differences lie that more outliers occur and the maximum outlier value becomes larger when the data's dimension is increased from $33$ to $272$, which makes it much easier to detect the anomaly behavior of the data. What's more, the $\mathcal{N}_\phi-t$ curve corresponding to high data dimension is smoother, which indicates the designed increasing data dimension algorithm can help improve the approach's robustness against random disturbance and measurement error.

\begin{figure}[!t]
\centerline{
\includegraphics[width=2.5in]{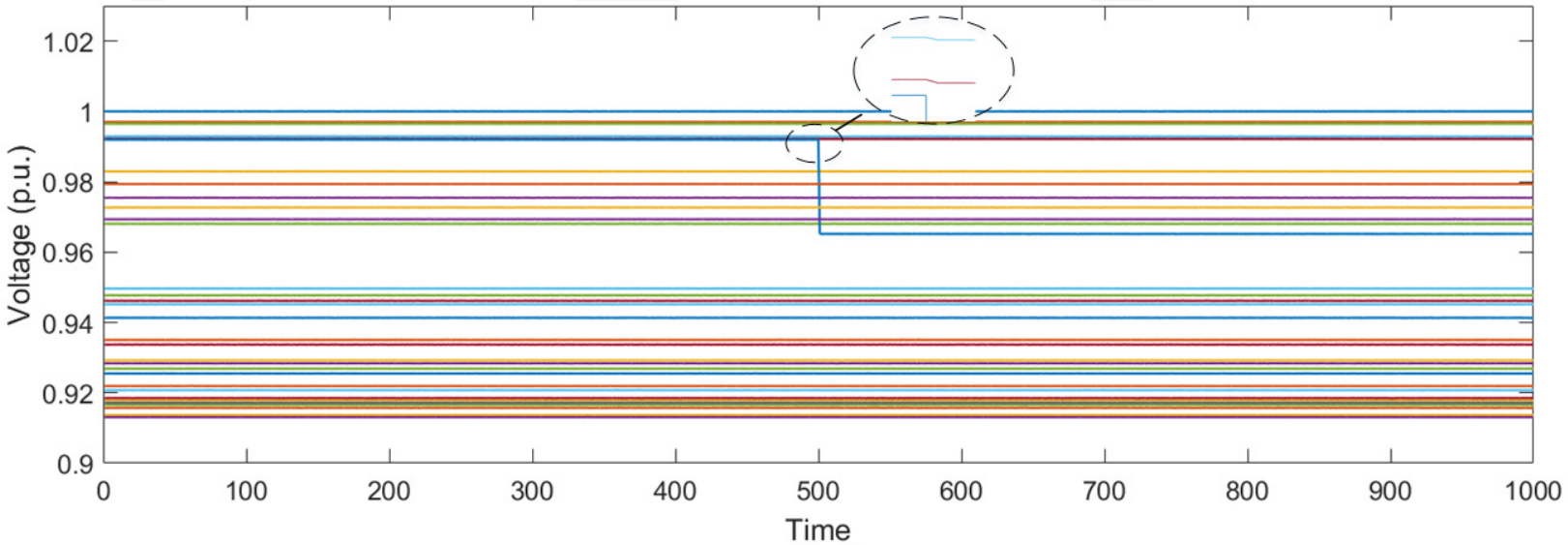}
}
\caption{{The voltage measurement curves in Case 3. The anomaly indexes were $20\sim 22$.}}
\label{fig:case3_data_org}
\end{figure}

3) Case Study On Anomaly {Localization}: In this case, the developed anomaly {localization} approach was tested by using the simulation data generated from IEEE 33-bus system. During the simulation, the anomaly signal was set the same as in Case 2, as shown in Table \ref{Tab: Case2}. The data consisted of $33$ voltage measurements for a series of $1000$ sampling times. The {voltage measurement curves} with anomaly indexes labelled were plotted in Figure \ref{fig:case3_data_org}. In our experiment, the the moving window's size was $33\times 200$, and the signal-to-noise ratio $\tau_{SNR}$ defined in Case 1 was $1000$.

\begin{figure}[!t]
\centering
\begin{minipage}{4.1cm}
\centerline{
\includegraphics[width=1.83in]{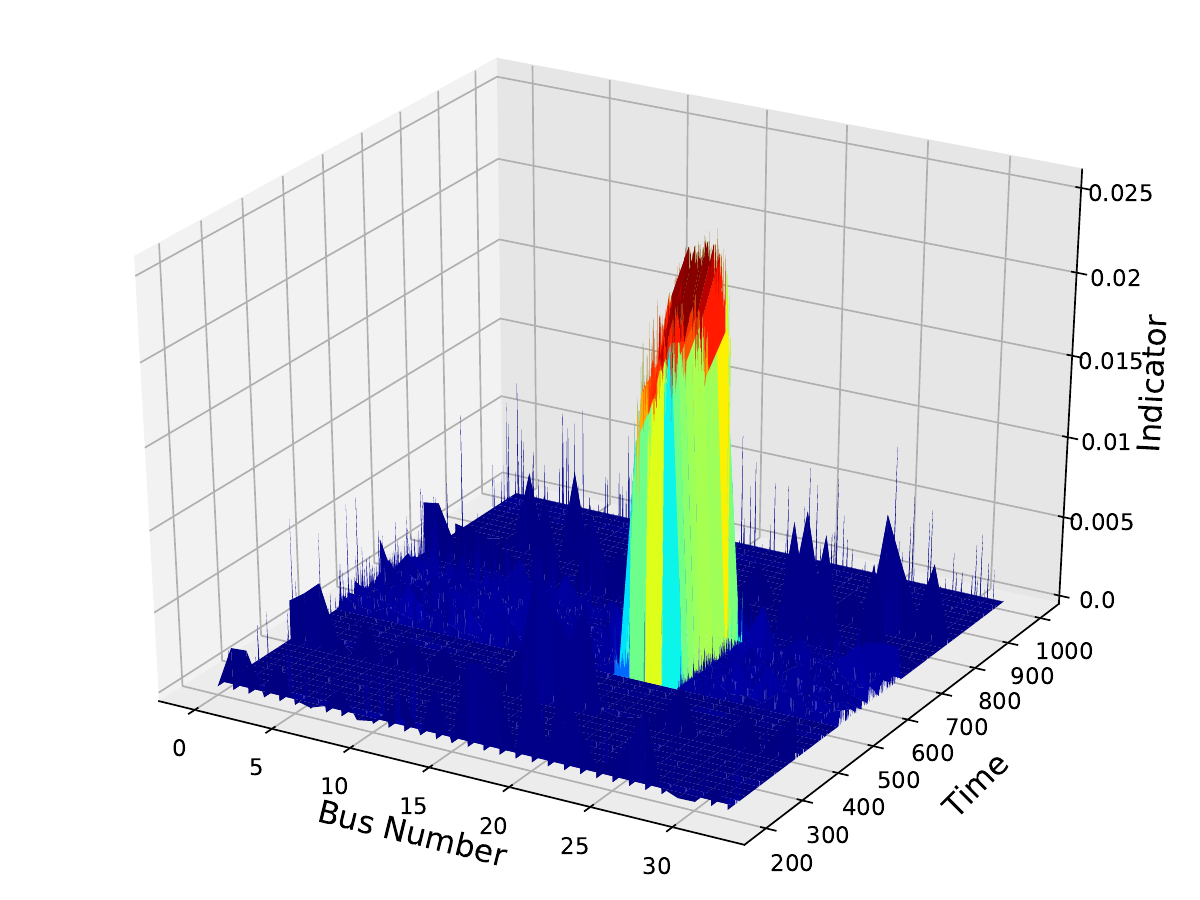}
}
\parbox{5cm}{\small \hspace{1.2cm}(a) Dimension: $33$}
\end{minipage}
\hspace{0.2cm}
\begin{minipage}{4.1cm}
\centerline{
\includegraphics[width=1.83in]{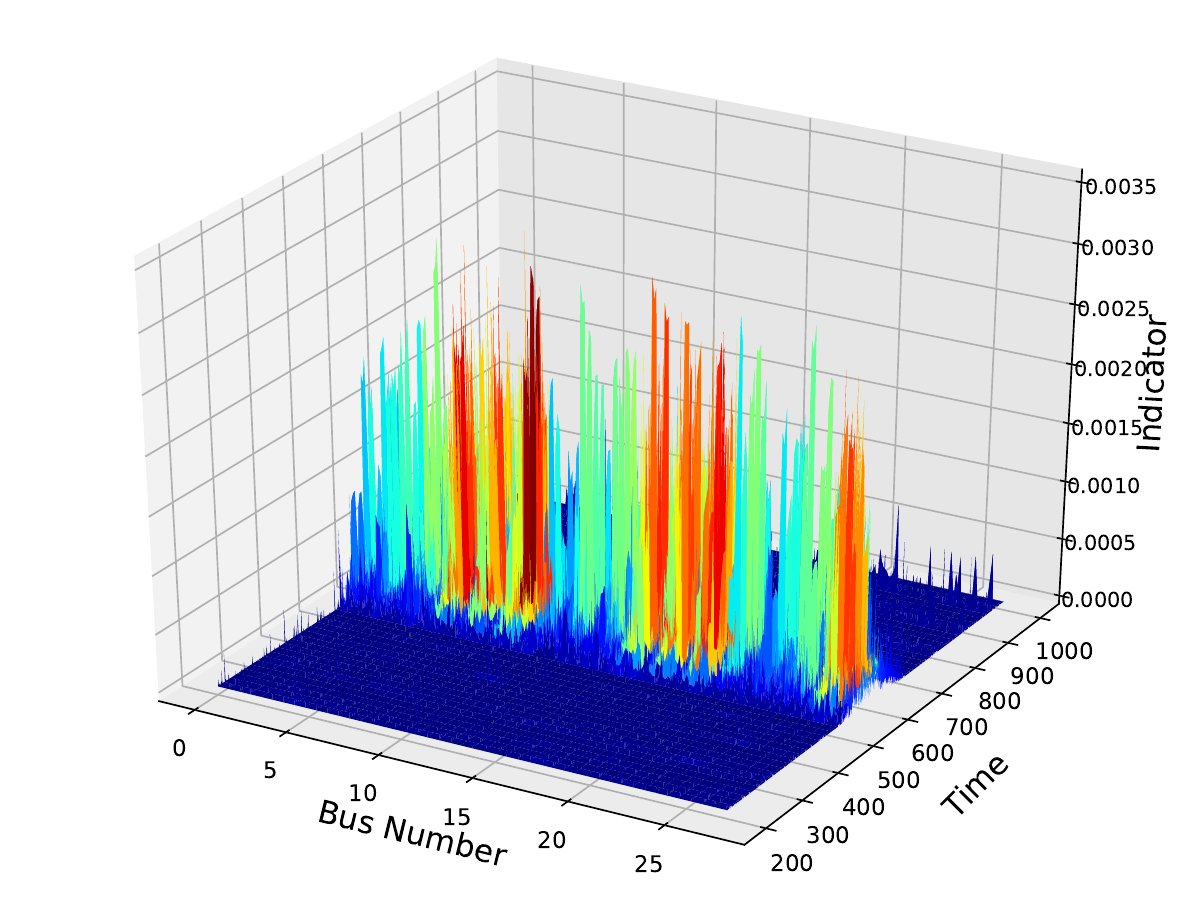}
}
\parbox{5cm}{\small \hspace{1.2cm}(b) Dimension: $272$}
\end{minipage}
\caption{The anomaly {localization} results corresponding to different data dimension. The dimension of the data was increased from $33$ to $272$.}
\label{fig:case3_loc}
\end{figure}

Figure \ref{fig:case3_loc}(a) plots the anomaly {localization} result. It can be seen that, from $t_s=501$, the {localization} indicator $\eta_{20\sim 22}$ increase dramaticlly and others stay almost unchanged, which denotes the anomaly occurs on $20\sim 22$ buses. Taking the sampling time $t_s=501$ as an example, the value of $1-\alpha$ for bus $20\sim 22$ and other buses (e.g., $23$) are $96.65\%$, $99.59\%$, $99.91\%$, and $24.38\%$, respectively. The {localization} result coincides with the recorded anomaly indexes.

Furthermore, we explore the effectiveness of the {localization} approach when the data's dimension is increased from $33=16+17$ to $272=16\times 17$. Figure \ref{fig:case3_loc}(b) plots the anomaly {localization} result. It can be seen that, from $t_s=501$, the {localization} indicator $\eta_{3\sim 5, 20\sim 22,\cdots,258\sim 260}$ increase dramaticlly and others stay almost unchanged. Let ${\bf I}=\{3\sim 5, 20\sim 22,\cdots,258\sim 260\}$, at $t_s=501$, the calculated $1-\alpha$ corresponding to $\bf I$ and others (such as $6$) are $\{98.84\%\sim 99.50\%, 86.04\%\sim 87.35\%,\cdots, 99.63\%\sim 99.77\%\}$ and $44.23\%$, respectively. Assume $n=17$, the anomaly location in the original data matrix can be  calculated by $\{({\bf I}\;{\bf{mod}}\;n)+n\}$, i.e., $\{20\sim 22\}$, which coincides with the real anomaly indexes.

\begin{figure}[!t]
\centerline{
\includegraphics[width=3.0in]{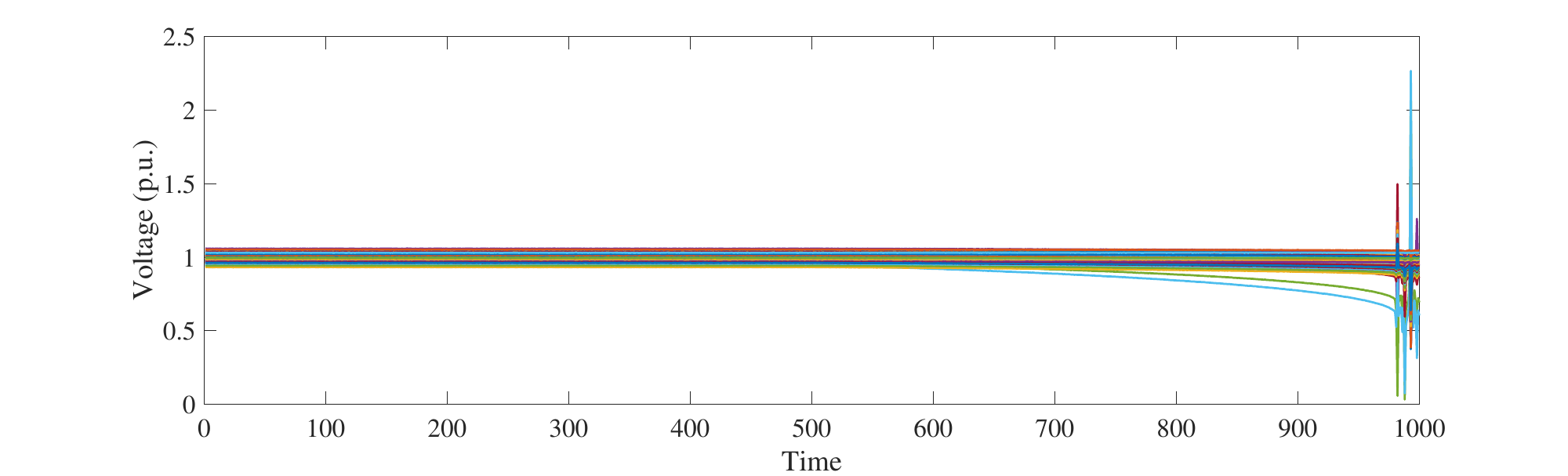}
}
\caption{The generated voltage measurement curves in Case 4.}
\label{fig:case4_data_org}
\end{figure}

4) Case Study on Comparison with Existing Techniques: In this case, we make a comparison on our {RMT based approach} with SVM \cite{ma2003time}, AE \cite{liu2018anomaly} and LSTM \cite{malhotra2015long} to illustrate the advantage of our approach. The IEEE 57-bus test system was used to {produce the simulation data}. In the simulation, an anomaly signal was set by increasing the active load of bus $20$ gradually, as illustrated in Table \ref{Tab: Case4}. The simulation data consisted of $57$ voltage measurements for a series of $1000$ sampling times, as plotted in Figure \ref{fig:case4_data_org}. In the experiment, the signal-to-noise ratio $\tau_{SNR}$ was set as $1000$. For SVM, AE or LSTM, a prediction model was firstly trained by using the normal data sequence during $t_s=1\sim 200$ and calculated the predicting error for the rest sequence during $t_s=201\sim 1000$, where {the prediction error was considered as the anomaly indicator and each sampling was used as a data sample. The experimental parameters involved in SVM, AE and LSTM can refer Table IV in our previous work \cite{shi2019spatio}.} For our approach, both the approach itself (RMT) and its combination with the designed increasing data dimension algorithm (RMT+IDD) were tested. In our approach, the moving window's size was set as $57\times 200$, the parameter $\tau_{\alpha}$ was $1$, and the data's dimension was increased from $57=28+29$ to $812=28\times 29$.
\begin{table}[!t]
\caption{The Anomaly Signal Set at Bus $20$ in Case 4.}
\label{Tab: Case4}
\centering
\footnotesize
\begin{tabular}{clc}   
\toprule[1.0pt]
\textbf{Bus} & \textbf{Sampling Time}& \textbf{Active Load(MW)}\\
\hline
\multirow{2}*{20} & $t_s=1\sim 500$ & $10$ \\
~&$t_s=501\sim 1000$ & $10\rightarrow 60$ \\
Others & $t_s=1\sim 1000$ & Unchanged \\
\hline
\end{tabular}
\end{table}

\begin{figure}[!t]
\centerline{
\includegraphics[width=3.0in]{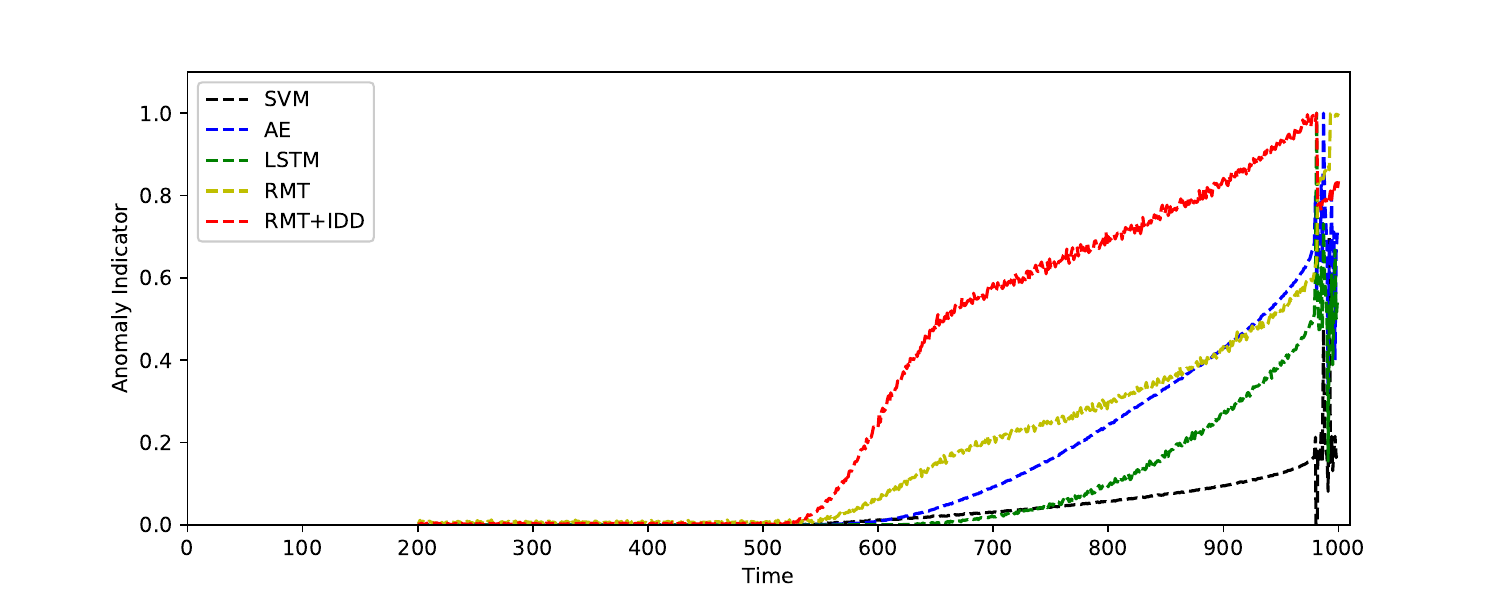}
}
\caption{The comparison result of different anomaly detection techniques in Case 4.}
\label{fig:case4_comparison}
\end{figure}
Figure \ref{fig:case4_comparison} shows the comparison result of different anomaly detection techniques. For SVM, the normalized value of the signed distance to the separating hyperplane was plotted; for AE and LSTM, the normalization results of the predicting errors were plotted; for RMT and RMT+IDD, $1-{\hat{\mathcal{N}}}(\phi)$ was plotted, where ${\hat{\mathcal{N}}}_\phi$ was the normalized form of $\mathcal{N}_\phi$. It can be observed that, {SA+IDD and SA are able to detect the anomaly signal at $t_s=530\sim 540$, which is much earlier than the other approaches ($t_s=610\sim 630$)}. It validates that our approach is more sensitive to the {abnormal data behavior} and it is capable of detecting the anomaly at an early stage. That's because a large moving data window rather than just the current data sample for each sampling time is analyzed in our approach. The average result makes it more robust against random disturbance and measurement error of the data. Moreover, it is noted that RMT+IDD outperforms RMT in anomaly detection, which indicates the designed increasing data dimension algorithm can help improve the sensitivity of RMT approach for abnormal data behavior.

Furthermore, the $average\; calculating\; time \; (ACT)$ for each sampling data  was counted to compare the efficiency of different detection techniques. For SVM, AE and LSTM, the $ACT$ for each data sample in the remaining sequence was calculated, and it did not consist of the model training time. In the experiments, the central processing unit of the computer server was 2.6 GHz, and the random access memory was 8 GB. The $ACT$ for SVM, AE, LSTM, RMT and RMT+IDD are $0.001$, $0.001$, $0.002$, $0.002$ and $0.029$ (unit: s), respectively. Considering the developed approach is an unsupervised approach without any training, it can be concluded that the approach has competitive performance in efficiency.
\subsection{Case Study with Real {SCADA Data}}
\label{subsection: case_real}
The {measurement data collected from the SCADA system installed in the distribution network} in Hangzhou city of China was used to verify the effectiveness of the developed approach. The distribution network consisted of $200$ feeder lines. For every feeder, multiple {measurement devices} were installed in different physical locations, through which the data was sampled at 15 minute intervals. The anomaly time and location index were recorded during the feeders'operation. In Case 5 and Case 6, three-phase voltage, three-phase current and active load sampled from March 1st, 2017 to March 14th, 2017 were used to form  the spatio-temporal data matrices.

\begin{figure}[!t]
\centering
\begin{minipage}{4.1cm}
\centerline{
\includegraphics[width=1.90in]{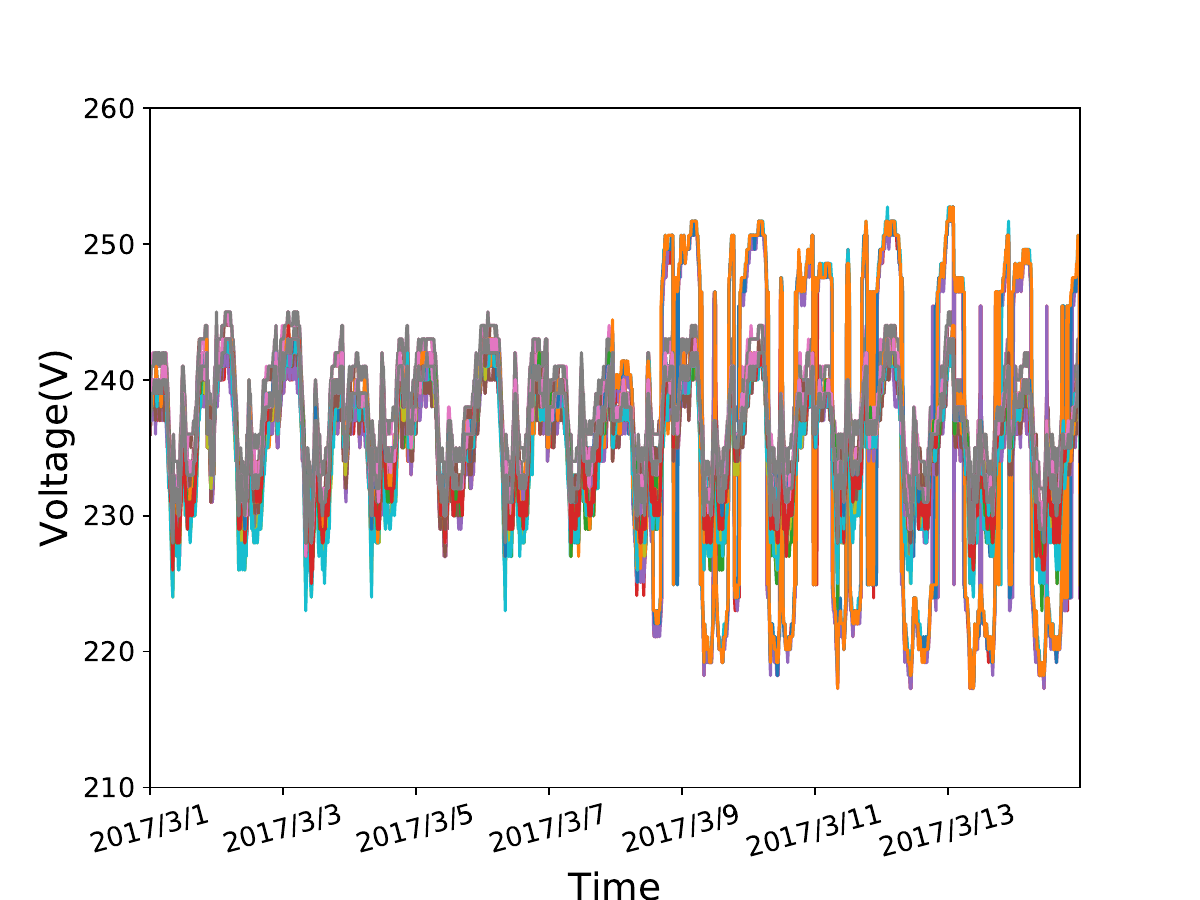}
}
\parbox{5cm}{\small \hspace{1.8cm}(a)}
\end{minipage}
\hspace{0.2cm}
\begin{minipage}{4.1cm}
\centerline{
\includegraphics[width=1.90in]{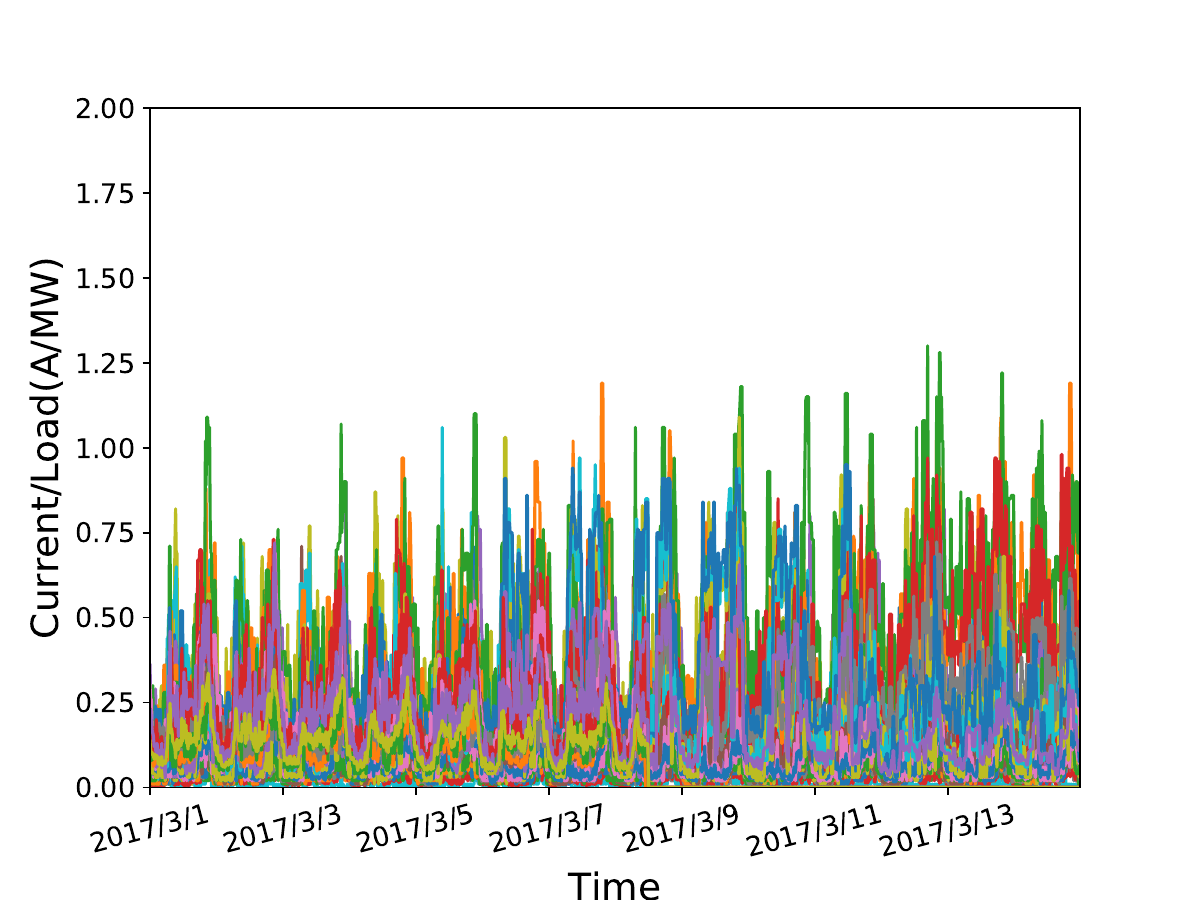}
}
\parbox{5cm}{\small \hspace{1.8cm}(b)}
\end{minipage}
\caption{The measurement data with anomaly time and location indexes recorded in Case 5. (a) Three-phase voltage magnitude curves. (b) Three-phase current and active load magnitude curves. The anomaly time was 2017/3/8 13:45:00 and the anomaly location indexes were $53, 54, 59, 60, 61$.}
\label{fig:case5_real_org}
\end{figure}
5) Case Study on High Observability Feeders: In this case, the developed approach was validated by analyzing a high-dimensional measurement data set from one feeder line. The data was collected from $17$ monitoring devices deployed on the feeder and it was consisted of $17\times 7=119$ measurement variables, thus a $119\times 1344$ data set was formed. The measurements with anomaly time and location recorded were plotted in Figure \ref{fig:case5_real_org}. The three-phase voltage curves indicated the anomaly was caused by voltage disturbance (violation). In the experiment, the moving window's size was set as $119\times 192$. The generated $\mathcal{N}_\phi-t$ curve with continuously moving windows was plotted in Figure \ref{fig:case5_detection}, in which the anomaly time was marked with a red dashed line. The early anomaly detection process is shown as follows:
\begin{figure}[!t]
\centerline{
\includegraphics[width=2.2in]{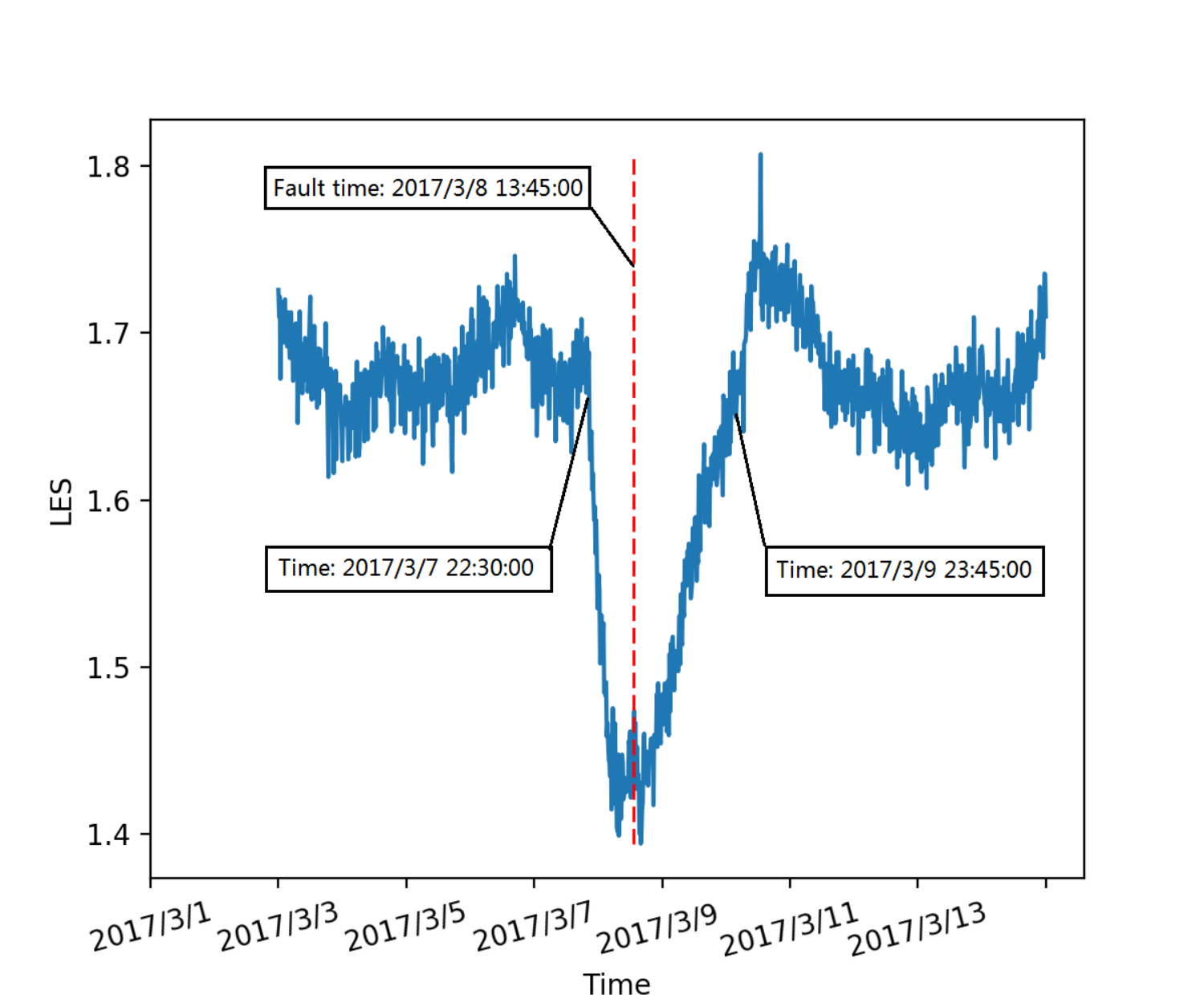}
}
\caption{The anomaly detection result in Case 5.}
\label{fig:case5_detection}
\end{figure}

\uppercase\expandafter{\romannumeral1}. During 2017/3/3 00:00:00$\sim$2017/3/7 22:30:00, $\mathcal{N}_\phi$ remains nearly constant, which denotes the feeder is in steady state.

\uppercase\expandafter{\romannumeral2}. From 2017/3/7 22:30:00, $\mathcal{N}_\phi$ begins to decrease dramaticlly, which denotes early anomaly signals occur and the operational state of the feeder begins to deteriorate. In view of the fact that the anomaly time is 2017/3/8 13:45:00, it can be concluded that the anomaly is detected in an early phase by the developed approach. Meanwhile, it is noted that, from 2017/3/7 22:30:00 to 2017/3/9 23:45:00, the $\mathcal{N}_\phi-t$ curve is almost $\bf U$ shaped and the duration of the anomaly signal on $\mathcal{N}_\phi$ is determined by the window's width, which coincides with our simulation result in Case 1.

\begin{figure}[!t]
\centerline{
\includegraphics[width=2.2in]{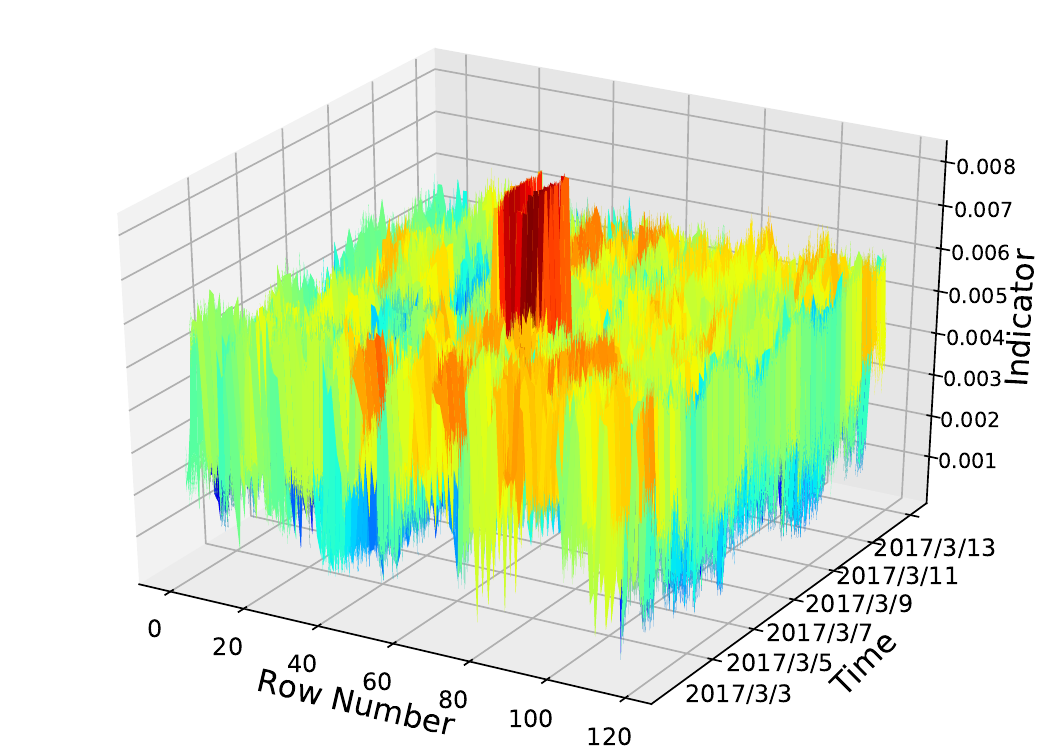}
}
\caption{The anomaly {localization} result in Case 5.}
\label{fig:case5_location}
\end{figure}
Figure \ref{fig:case5_location} is the 3D plot of the {localization} indicator $\eta$ regarding $119$ measurement variables from 2017/3/3 00:00:00 to 2017/3/14 23:45:00. It can be observed that, from 2017/3/7 22:30:00, $\eta_{\{53, 54, 59, 60, 61\}}$ increase rapidly and they are larger than others (such as $\eta_{100}$), thus the anomaly index set is determined as ${\bf I}=\{53, 54, 59, 60, 61\}$. For example, at 2017/3/7 22:45:00, the calculated values of $1-\alpha$ corresponding to $\eta_{\bf I}$ and $\eta_{100}$ are $99.89\%,99.91\%,99.92\%,99.91\%,99.93\%$ and $34.65\%$, respectively. The {localization} result coincides with the recorded anomaly indexes.

\begin{figure}[!t]
\centering
\begin{minipage}{4.1cm}
\centerline{
\includegraphics[width=1.90in]{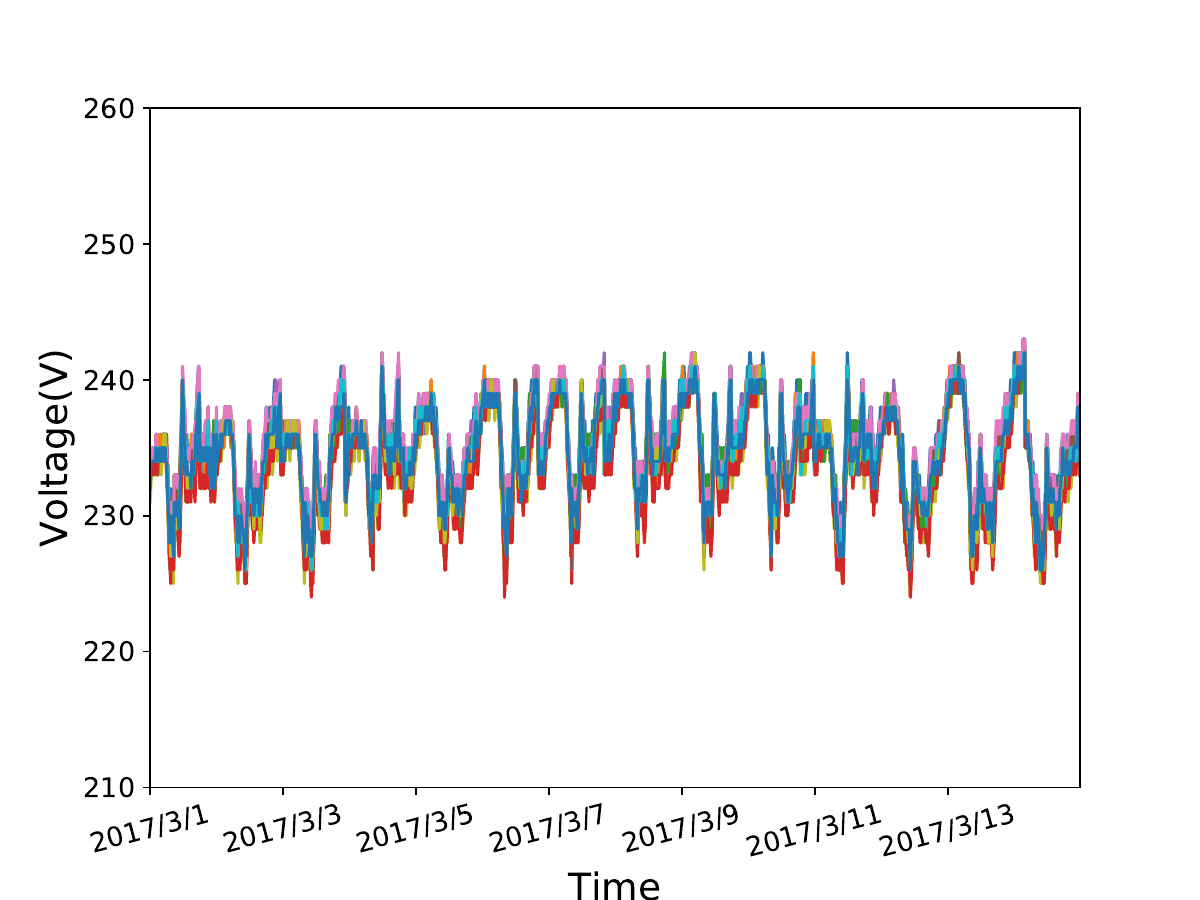}
}
\parbox{5cm}{\small \hspace{1.8cm}(a)}
\end{minipage}
\hspace{0.2cm}
\begin{minipage}{4.1cm}
\centerline{
\includegraphics[width=1.90in]{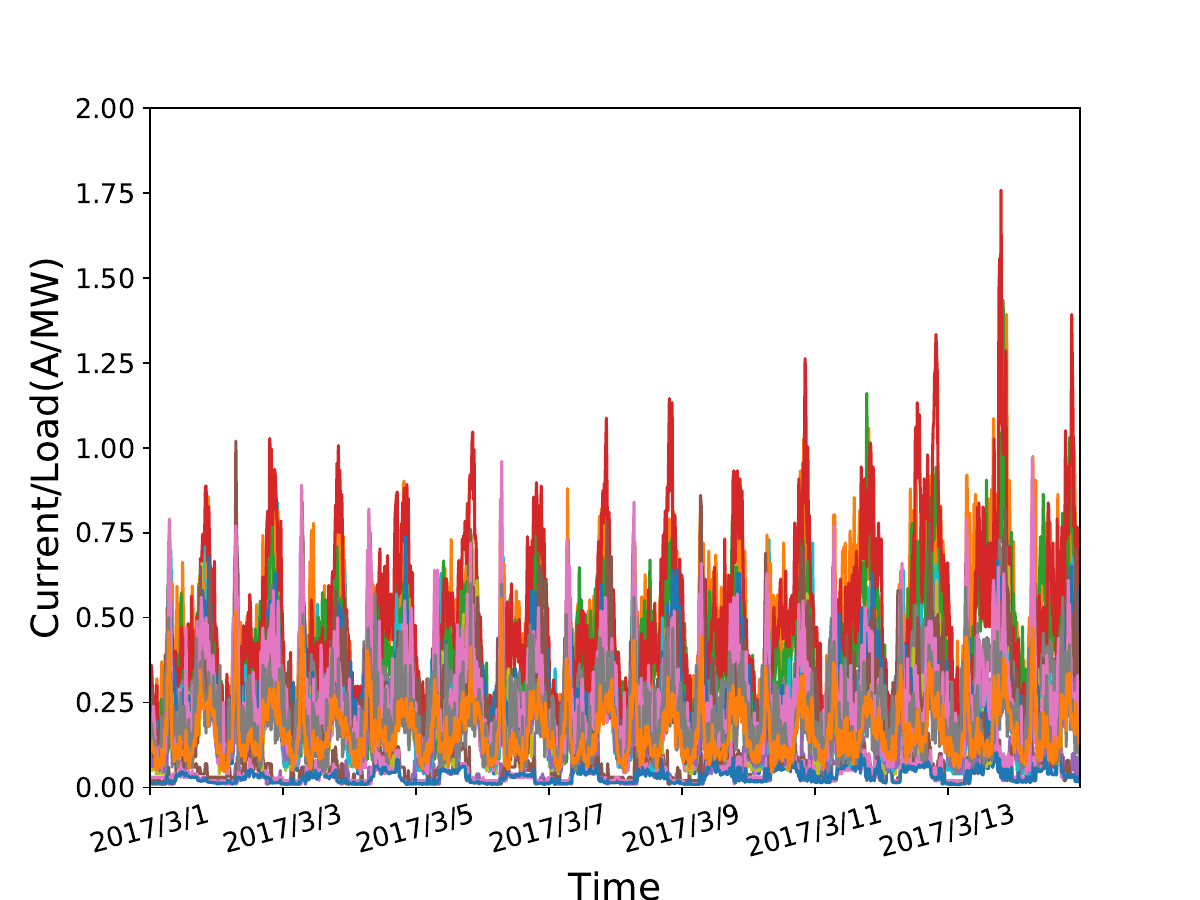}
}
\parbox{5cm}{\small \hspace{1.8cm}(b)}
\end{minipage}
\caption{The measurement data with anomaly time and location indexes recorded in Case 6. (a) Three-phase voltage magnitude curves. (b) Three-phase current and active load magnitude curves. The anomaly time and indexes were 2017/3/8 13:30:00, 2017/3/13 03:00:00 and $\{1,2,3\}$, $\{1,2,7\}$, respectively.}
\label{fig:case6_real_org}
\end{figure}
6) Case Study on Low Observability Feeders: In this case, the combination of the designed increasing data dimension algorithm and the developed approach was verified by using a low-dimensional {measurement} data set from one low observability feeder line. The data was collected from $6$ monitoring devices and it was consisted of $6\times 7=42$ measurement variables, thus a $42\times 1344$ data set was formed. The measurements with anomaly time and location indexes recorded were plotted in Figure \ref{fig:case6_real_org}. The three-phase current and active load curves demonstrated that the anomaly was caused by overload. In the experiment, the moving window's was set as $42\times 192$. For each moving data window, the dimension was increased from $42=21\times 2$ to $21^2=441$ through the proposed increasing data dimension algorithm, in which $\tau_{\alpha}$ was set as $1$. The generated $\mathcal{N}_\phi-t$ curve with continuously moving windows was plotted in Figure \ref{fig:case6_detection}, in which the anomaly time was marked with red dashed lines. The early anomaly detection process is shown as follows:
\begin{figure}[!t]
\centerline{
\includegraphics[width=2.2in]{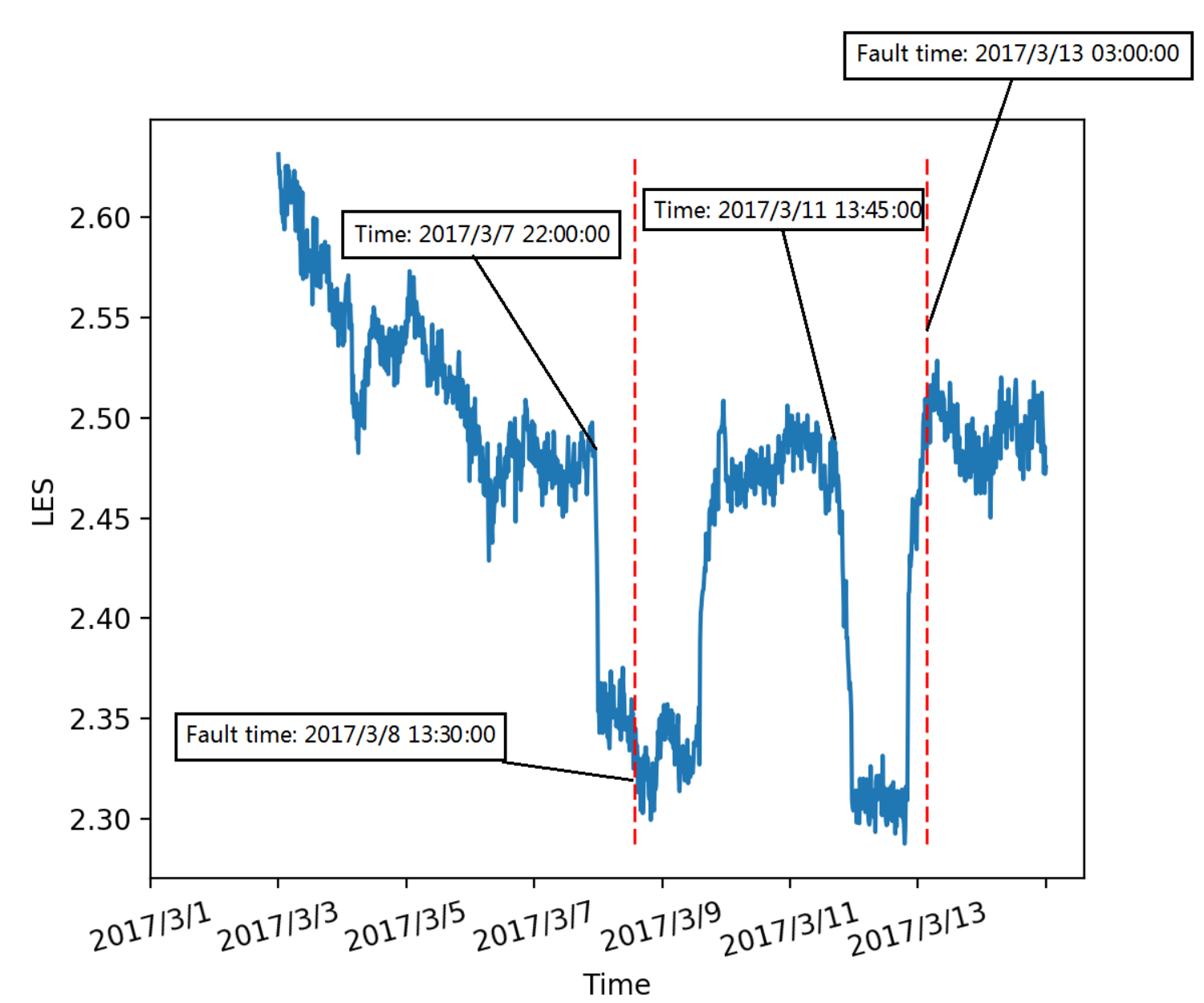}
}
\caption{The anomaly detection result in Case 6.}
\label{fig:case6_detection}
\end{figure}

\uppercase\expandafter{\romannumeral1}. During 2017/3/3 00:00:00$\sim$2017/3/7 22:00:00, $\mathcal{N}_\phi$ decreases gradually, which indicates the operating state of the feeder is getting worse and the anomaly may occur at any time.

\uppercase\expandafter{\romannumeral2}. From 2017/3/7 22:00:00, $\mathcal{N}_\phi$ decreases dramatically, which denotes an anomaly signal occurs and the feeder is in {unsteady} state. In view of the fact that  the anomaly time is 2017/3/8 13:30:00, it can be concluded that the anomaly is detected in an early phase. Meanwhile, it is noted that, from 2017/3/7 22:00:00 to 2017/3/9 22:00:00, the $\mathcal{N}_\phi-t$ curve is almost $\bf U$-shaped and the duration of the anomaly signal on $\mathcal{N}_\phi$ is determined by the moving window's width, which coincides with the simulation result in Case 1. Similarly, from 2017/3/11 13:45:00, another new anomaly signal is detected, which is much earlier than the recorded anomaly time.

\begin{figure}[!t]
\centerline{
\includegraphics[width=2.2in]{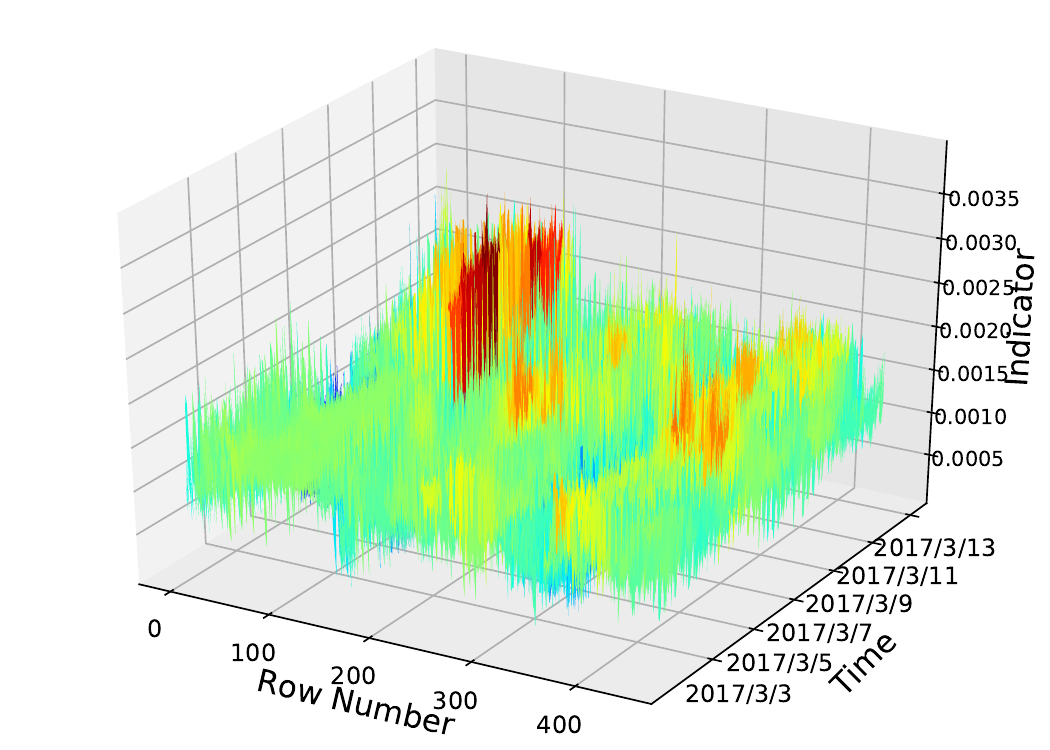}
}
\caption{The anomaly {localization } result in Case 6.}
\label{fig:case6_location}
\end{figure}
Furthermore, we {localized} the anomalies through the developed approach, {as} shown in Figure \ref{fig:case6_location}. It can be observed that, from 2017/3/7 22:00:00, $\eta_{1\sim 63}$ increase rapidly and they are larger than others (such as $\eta_{100}$), which indicates the anomaly index set ${\bf I}=\{1\sim 63\}$. For example, at 2017/3/7 22:15:00, the values of $1-\alpha$ corresponding to $\eta_{\bf I}$ and $\eta_{100}$ are $96.74\%\sim 98.91\%$ and $42.32\%$, respectively. Let $n=21$, then the anomaly indexes in the original data set can be calculated by $\{{\bf I}\;{\bf{mod}}\;n\}$, i.e., $\{{1,2,3}\}$. Similarly, from 2017/3/11 13:45:00, $\eta_{1\sim 42,106\sim 126}$ increase rapidly and they are larger than others (such as $\eta_{200}$), which determines the anomaly index set ${\bf I'}=\{1\sim 42,106\sim 126\}$. For example, at 2017/3/11 14:00:00, the calculated values of $1-\alpha$ corresponding to $\eta_{\bf I'}$ and $\eta_{200}$ are $99.97\%\sim 99.99\%, 99.99\%\sim 99.99\%$ and $38.27\%$, respectively. Then the anomaly indexes in the original data set can be calculated by $\{{\bf I'}\;{\bf{mod}}\;n\}$, i.e., $\{{1,2,7}\}$. The {localization} results coincide with the real anomaly indexes.
\section{Conclusion}
\label{section: conclusion}
Based on the RMT, a data-driven approach is developed for early anomaly detection and {localization} in distribution network. It is able to detect and {localize} the anomaly at an early stage by tracking the variation of the data correlations. The linear eigenvalue statistics give insight into the data behavior from a high-dimensional perspective, which is used as the detection indicator in the developed approach. As for the low observability feeders in the distribution network, an increasing data dimension algorithm is designed for them to be analyzed more accurately. The developed approach is mainly data-driven without requiring complex parameter information of the distribution network. It merges anomaly detection and {localization} functionalities, and is robust against random disturbance and measurement error. Case studies on the simulation data and the real {SCADA} data corroborate the feasibility and advantages of the approach.

In our future work, we will focus on two aspects: {1) realizing anomaly declare automatically. Specifically, we will explore an indicator based on the $LES-t$ curve to measure the degree of an anomaly by comparing it with the pre-defined threshold. The optimal threshold value should be searched by a designed algorithm so that the developed approach has a higher detection accuracy rate and a lower false alarming rate for a given data set}. 2) realizing the analysis of different types of faults, such as single-phase fault, two-phase fault, three-phase fault, etc. The linear eigenvalue statistics via different test functions are considered as different filters, and we can use them to track different types of fault signals.

\small{}
\bibliographystyle{IEEEtran}
\bibliography{helx}

\normalsize{}
\end{document}